\documentclass[11pt,a4paper]{article}

\usepackage[utf8]{inputenc}
\usepackage[T1]{fontenc}
\usepackage{amsmath,amssymb,amsfonts}
\usepackage{graphicx}
\usepackage{hyperref}
\usepackage{natbib}
\usepackage{booktabs}
\usepackage{float}
\usepackage{dsfont}
\usepackage{geometry}
\title{Mapping Urban Air Quality from Mobile Sensors Using Spatio-Temporal Geostatistics}

\author{
Yacine Mohamed Idir$^{1,2,3}$, Olivier Orfila$^{1}$, Vincent Judalet$^{3}$, Benoit Sagot$^{3}$, Patrice Chatellier$^{2}$\\[0.5em]
\small $^{1}$ COSYS-PICS-L, Univ Gustave Eiffel, IFSTTAR, F-78000 Versailles, France\\
\small $^{2}$ COSYS-LISIS, Univ Gustave Eiffel, IFSTTAR, F-77454 Marne-la-Vallée, France\\
\small $^{3}$ ESTACA, F-78066 Saint-Quentin-en-Yvelines, France\\[0.5em]
\small Correspondence: mohamed-yacine.idir@univ-eiffel.fr
}

\date{}

\begin{document}

\maketitle

\begin{abstract}
With the advancement of technology and the arrival of miniaturized environmental sensors that offers greater performance, the idea of building mobile network sensing for air quality has quickly emerged to increase our knowledge of air pollution in urban environments. However, with these new techniques, arise the difficulty to build mathematical models capable of aggregating all these data sources in order to provide a precise mapping of air quality.
In this context, this paper explores the spatio-temporal geostatistic methods as a solution for such a problem and evaluates three different methods: simple kriging in residuals, ordinary kriging, and kriging with external drift.
We conclude that univariable geostatistics are suitable for interpolating this type of data but are less performant for an extrapolation of non-sampled places, since it does not create any information.
\end{abstract}


\section{Introduction}

Air pollution is one of the major concerns of the last century, causing more than 7 million deaths per year \cite{WHO}. The situation is more alarming in metropolitan areas where air quality regularly exceeds the standards suggested by the World Health Organization \cite{sharma2013response}. This can be attributed to the scale of urbanization and population growth, as well as the resulting energy consumption \cite{manisalidis2020environmental}.
Air quality monitoring is a crucial part in the process of reducing urban air pollution or at least its harmful effects on people's health and the environment. Indeed, real-time information on air pollution in urban areas is of great importance for environmental and health protection agencies which must advise the general public as soon as possible. These information can also be used by companies to offer several services and solutions in order to reduce the impact of air pollution on health.

\subsection{Fixed Pollution Sensors}

Currently, air quality monitoring is carried out using fixed air quality monitoring stations. These stations are managed by national environmental protection agencies like the Automatic Urban and Rural Network (AURN)\footnote{\url{https://uk-air.defra.gov.uk/networks/network-info?view=aurn}} in the United Kingdom, and AirParif\footnote{\url{https://airparif.asso.fr}} in France. These reference stations provide a very precise measurement of air quality at the cost of limited spatial coverage.
The stations can generate detailed time series data (usually with hourly resolution), but at limited locations. This makes it difficult to compile reliable and representative information for a city or a region as a whole and, therefore, a more macroscopic view of trends in pollution fields is provided.

However, the air quality in a city varies a lot, because the concentration of pollutants in a given place depends mainly on local emission sources and atmospheric flow conditions \cite{britter2003flow}. The flow of air masses in urban environments is generally turbulent and difficult to predict without sophisticated numerical modelling tools.
For example, after comparing surveillance data from two streets in Copenhagen (\textit{Jagtvej} and \textit{Bredgade}), Berkowicz et al. \cite{berkowicz1996using} argued that roadside sightings are site dependent and not representative of a larger urban area. They demonstrated that the measured concentrations could be very different at these two sites, mainly due to the different positions of the monitoring stations in the streets. Another study \cite{scaperdas1999assessing}, showed that the air quality measurements taken at the intersection of two central London streets were highly dependent on local wind flow and the geometry of the streets and buildings surrounding the receiver.

Knowing that the total number of fixed air quality monitoring stations in a city is limited due to practical constraints, such as the cost and size of equipment, the power supply, and so on. Thus, the increase in the number of fixed stations is often hard to achieve. Hence, it is necessary to use other measurement and modelling techniques to assess urban air quality at unsampled places.

There exist five large families of models and methods for creating urban air pollution cartography using fixed sensors:

\textbf{Land-Use Regression models.} Land-Use Regression models (LUR) make the assumption that the air quality in a given place depends only on the local characteristics of the environment, such as land use, weather-related variables, building density, traffic density, and so on. These models link the measurement of air quality taken at the fixed station to the chosen predictive environmental variables.
A LUR model developed by Kerckhoffs et al. \cite{kerckhoffs2015national}, including small-scale traffic, large-scale address density and urban green, explained 71\% of the spatial variation in summer average ozone concentrations. Meng et al. \cite{meng2015land} and Chen et al. \cite{chen2010land} successfully developed a LUR model for NO2 concentrations in China. Another example of LUR model was built in Italy to assess NO2 concentrations \cite{marcon2015development}.
LUR models provide good results for a rather low complexity. They also describe the effect of the environmental variables on the pollutant concentration, but remains limited by the amount of data from other variables needed or obtained at a relatively expensive cost.

\textbf{Deterministic interpolation methods.} One of the most popular deterministic interpolation methods is the Inverse Distance Weighting (IDW). The value at the unknown location is calculated as the weighted average of the measurements collected from the monitoring stations. This method assumes that the value is more influenced by the nearest measurements than the distant ones, the closest locations obtain greater weights. As the distance increases, less weight is given to the measurement.
Given the simplicity of this method, it is often used as a benchmark. Marshall et al. \cite{marshall2008within} used it to compare the urban variability of NO and NO2 concentration to a LUR model and an Eulerian grid modelling in Vancouver, Canada. Wong et al. \cite{wong2004comparison} compare different interpolation methods including IDW to estimate O3 and PM10 air concentrations.
The weakness of deterministic interpolation methods lies in their poor extrapolation accuracy. These methods are not considered as models, because they do not describe the data in addition to not giving uncertainty associated with the prediction.

\textbf{Geostatistics.} Geostatistics regroup stochastic kriging methods, the value at the unsampled location is evaluated by a weighted linear combination of measurements, the weights are calculated from the variability of the data inferred from the actual spatial structure of the data.
Kim et al. \cite{kim2014ordinary} developed ordinary kriging prediction models to predict long-term particulate matter concentrations in seven major Korean cities. Whitworth et al. \cite{whitworth2011kriged} modelled ambient air levels of benzene in an urban environment. More sophisticated than IDW and regression modelling, geostatistics also provide the uncertainty associated with the prediction. However, these techniques suffer from relatively a high computational cost.

\textbf{Dispersion models.} Dispersion models replicate the formation of atmospheric pollutants through physical and chemical processes. They have been widely used in traffic related pollution prediction with making use of the environmental variables such as the ones used in LUR models.
Hamer et al. \cite{hamer2019urban} describe the Eulerian urban dispersion model EPISODE and its application to the modelling of NO2 pollution concentration. Fallah et al. \cite{fallah2017integrating} improved characterisation of near-road air pollution using regional Gaussian dispersion model. Gibson et al. \cite{gibson2013dispersion} used AERMOD Gaussian plume air dispersion model to evaluate the PM2.5, NOx and SO2.
However, these methods suffer from numerous shortcoming such as the computational cost and the production of uniform and imprecise maps, related to the challenging task of modelling the small scale random variations.

\textbf{Machine learning algorithms.} A machine learning algorithm analyses the training data and produces an inferred function, which can be used to map new examples. Machine Learning is very effective in situations where insights must be discovered from large sets of diverse and changing data. 
Numerous studies applied this method to predict air pollution levels: Singh et al. \cite{singh2013identifying} identify pollution sources and predict urban air quality using ensemble learning methods. Cabaneros et al. \cite{cabaneros2019review} give a review of artificial neural network models for ambient air pollution prediction.
Machine learning algorithms are considered as black boxes with poor descriptive power and struggle to provide better results than the other models with limited data.

With recent technological advances, the proliferation of air quality low-cost sensors offers additional tools to refine the spatial-temporal characterization of air pollution levels \cite{morawska2018applications}. Numerous instruments from business entities, non-profits and startups have entered the market so far \cite{borghi2017miniaturized}.
The performance of these sensors can differ significantly between different models as well as between units of the same model, as indicated by field and laboratory evaluations \cite{feinberg2018long}.

Although having many advantages, the use of this new type of sensors to assess urban atmospheric pollution also presents inconveniences. Mainly, taken separately, the data from these sensors are often noisy and not very precise.
Studies \cite{munir2019analysing,johnson2018field} analysed the performance of low-cost air quality sensors as well as their benefits and their viability for monitoring air pollution levels in urban areas. None of the sensors tested showed good correlation with reference data in low ambient concentrations (0 to 15 $\mu$g/m$^3$ range).
Nonetheless deployed in large quantities and using the right calibration and prediction models, they are able to provide complex and complementary information to the fixed monitoring station.

\subsection{Mobile Sensors}

The use of a fleet of low-cost sensors on board of vehicles (cars, buses, trams, and so on) traveling in an urban area in order to have a better representation of pollutants is increasingly popular. As opposed to the traditional air quality monitoring stations, the use of a low-cost mobile sensor network that can dynamically travel through the environment will deliver data with unprecedented resolution \cite{devarakonda2013real,re2014urban}.
Some notable examples of research projects using low-cost sensors for monitoring air pollution include: the ``OpenSense'' projects in Switzerland \cite{hasenfratz2015deriving}, ``Array of Things'' in Chicago, United States \cite{catlett2017array}, the Imperial County Community Air Monitoring Network \cite{english2017imperial} in California, United States, ``Gotcha'' II in Shenzhen, China \cite{inproceedings} and ``Air Map Korea Project'' in major cities of south Korea.

In this context, a mobile sensor could be a good compromise between temporal resolution and spatial resolution, allowing high spatial cover over large areas without using a large number of fixed sensors. However, due to the reduced temporal resolution of any sampled location, it is challenging to generate pollution maps with high temporal resolution at daily or hourly time scales.

Air quality monitoring using mobile sensors is attracting an increasingly growing interest \cite{merbitz2012mobile,van2015mobile}. Several devices were developed to monitor in real-time the spatial and temporal variability of air quality using different instruments, technologies, and platforms. Gozzi et al. \cite{gozzi2016mobile} summarize the status of mobile monitoring of particulate matter.
Most of these studies used mobile monitoring to assess air pollution exposure or to study spatial and temporal characteristics. Only few studies were interested in producing urban map pollution using mobile monitoring at fine spatial-temporal scale.

Real-time information on a very localized scale are very important for the citizens to make decisions on their day-to-day activities like the best path to go from a point A to a point B, or the perfect time to go out to minimize the effects of air pollution on their health.

A range of methods exist to go beyond the spatial and temporal coverage of the mobile measurements and draw pollution maps. Studies naturally applied the same methods used for fixed stations to the new problem generated by the use of mobile sensors. Table~\ref{surveys} summarizes the main recent studies using mobile monitoring to map air pollution levels.

\begin{table}[H]
\centering
\caption{Mapping air quality studies using mobile sensors.}
\label{surveys}
\begin{tabular}{ccccc}
\toprule
\textbf{Article} & \textbf{Method} & \textbf{Area} & \textbf{Pollutant} & \textbf{Sensor carrier} \\
\midrule
Marjovi et al. \cite{marjovi2017extending} & LUR, machine learning (ANN) & Lausanne, Switzerland & UFP & Bus \\
Hart et al. \cite{hart2020monitoring} & LUR & Texas, USA & PM2.5 & Bike \\
Apte et al. \cite{apte2017high} & Reduction algorithm & Oakland, USA & NO, NO2, BC & Car \\
Hasenfratz et al. \cite{hasenfratz2014pushing} & LUR & Zurich, Switzerland & UFP & Tram \\
Hasenfratz et al. \cite{hasenfratz2015deriving} & LUR & Zurich, Switzerland & UFP & Tram \\
Marjovi et al. \cite{marjovi2015high} & LUR, probabilistic Graphical Model & Lausanne, Switzerland & UFP & Bus \\
Li et al. \cite{li2014estimating} & Kriging & Zurich, Switzerland & UFP & Tram \\
Lim et al. \cite{lim2019mapping} & LUR, machine learning & Seoul, South Korea & PM2.5 & Pedestrian \\
Adams et al. \cite{adams2016mapping} & ANN & Hamilton, Canada & NO2 & Van \\
Hankey et al. \cite{hankey2015land} & LUR & Minneapolis, USA & BC PM2.5 & Bike \\
Gressent et al. \cite{gressent2020data} & Kriging & Nantes, France & PM10 & Car \\
Do et al. \cite{do2020graph} & Autoencoders & Antwerp, Belgium & Several pollutants & Bike \\
Zhang et al. \cite{zhang2020real} & Machine learning & Songdo, Korea & CO2, PM2.5, PM10 & Car \\
Song et al. \cite{song2020deep} & Machine learning & Beijing, China & PM2.5 & Car \\
Van et al. \cite{van2020development} & LUR & Ghent, Belgium & BC & Bike \\
Guan et al. \cite{guan2020fine} & LUR, kriging & Oakland, California & NO2 & Car \\
Mariano et al. \cite{mariano2020pollution} & Decision trees & Zurich, Switzerland & UFP & Tram \\
Ma et al. \cite{ma2020fine} & Machine learning & China & PM2.5 & Car \\
\bottomrule
\end{tabular}
\end{table}

Land-Use Regression models have become the standard method. Hatzopoulou et al. \cite{hatzopoulou2017robustness} and Kerckhoffs et al. \cite{kerckhoffs2017robustness} have evaluated the robustness of LUR models developed from mobile air pollutant measurements and have concluded that mobile monitoring provided robust LUR models for predicting ultrafine particles concentrations. This partially explains the popular use of these models in mobile monitoring.

All the studies in Table~\ref{surveys} have proposed models that share the same weaknesses with the LUR models: they require (and are mainly based) on information provided by external variables. These variables are introduced into the model for investigating the link with the pollutant level, and the predicted pollutant value at unsampled locations is therefore derived from the knowledge of these variables at those locations.
In addition to being able to predict only at the locations sampled by these covariates, the difficulty of their acquisition as well as the additional computational cost represent real obstacles to the use of these methods. Moreover they have the disadvantage of producing maps with relatively large spatial and temporal resolutions. Actually, the final resolution of the prediction highly depends on the resolution of the covariates.

The problem worsens when we are interested in real time prediction. Either these covariates are sometimes available only after a given period of time, which makes them unavailable for real time prediction, either we use the predictions of these variables which can introduce a lot of uncertainties in the final result.

Geostatistics have the advantage of being able to incorporate covariates (Kriging with External Drift ``KED'', cokriging), but can also do without it (simple kriging, ordinary kriging), thus represent with the deterministic methods a way to produce maps without using other variables.
It has the advantage, compared to the deterministic interpolation methods, to give the variance associated with the prediction. Although geostatistics make stronger assumptions on the data. This model family was selected to tackle the real-time prediction problem because of all previously introduced advantages.

Few studies used geostatistics as a way to map air pollution using low cost mobile sensors, where Li et al. \cite{li2014estimating} and Guan et al. \cite{guan2020fine} on top of using several covariates in their geostatistical model, they used a likelihood based method making stricter assumption on the underlying distribution of the data and increasing the computational resources, making it challenging to use in real-time applications.
Gressent et al. \cite{gressent2020data} used, as opposed to likelihood method, a variogram based method. They chose a purely spatial model which does not take into account the temporal correlation of the data.

This paper aims to show the prediction efficiency of variogram based spatio-temporal geostatistics in the mapping process of air quality using mobile sensors without the use of external variables other than pollution data for real-time prediction purpose.

\section{Materials and Methods}

\subsection{Methodology}

As stated in the introduction above, there is a need for a method to generate real-time air pollution maps. In this section, the methodology used to assess the efficiency of spatio-temporal geostatistics is presented, by comparing different geostatistics models and show the potential gain compared to a standard inverse distance weighting which is the most common and known in practice. First the research question has been defined:

\emph{What are the best models of space-time geostatistics for predicting urban air pollution using mobile sensors and what are the benefits compared to a standard deterministic approach?}

To answer this question, different geostatistics models were compared to a reference method (IDW) to estimate pollution maps using a database from the OpenSense project. Comparison has been made on the basis of a selection of performance indicators. The remainder of this section develops each of these steps.

\subsubsection{Data}

Considering the limited number of studies carried out on urban air pollution with mobile sensors, the number of public datasets is limited. This paper uses the data from the OpenSense project to answer the research question (the detailed definition of the dataset is introduced in Subsection~\ref{sec:data}). Ozone concentration has been selected as the first pollutant to be examined in this study and the methodology remains the same for any other pollutant categories.

\subsubsection{Model Selection}

Three geostatistical approaches have been applied. Apart from mobile sensors data, two of them use fixed station data to predict air quality. Each of these three methods make different assumptions, which will be discussed in details in the theoretical Section~\ref{sec:theory}:

\begin{itemize}
\item Simple kriging with a varying known mean: the time series of the fixed monitoring station was chosen to be the overall mean.
\item Ordinary kriging with a constant piecewise mean but unknown.
\item Kriging with external drift: the data from the fixed monitoring station was used to estimate the underlying mean.
\end{itemize}

The originality of the proposed models lies in their capacity to rely on a variographic study to describe spatiotemporal variance.

\subsubsection{Variographic Study}

In this paper only the estimate of the variogram and not the covariance function was made, making less restrictive assumption on the stationarity of the random field.
In the calculation of the experimental variogram, Arnaud et al. \cite{arnaud2000estimation} recommend taking into account distances up to the half of the maximum distance encountered between two points of the field. Beyond that, the number of pairs of points involved in the calculation of the variogram decreases and reduces its robustness.
Knowing that the maximum distance between 2 points in this study is 12.8 km, variograms are thus limited to 6 km.
As for the temporal limit, knowing that months of data are available, restricting this study to half of this temporal distance is neither possible in practice nor advantageous. The retained limit is set manually by increasing the time limit step by step until a sill appears in the variogram.

One week of data was used to estimate the empirical variogram and the weighted least square technique from Cressie \cite{cressie1985fitting} to adjust a theoretical spatio temporal variogram to the sample variogram was chosen.
To study a possible anisotropy in the data linked to external factors, two spatio-temporal empirical variograms in the two static directions (North-South and East-West) were performed.
Finally, three variograms were computed, each one associated with a different selected model.

\subsubsection{Models Validation Process}

In order to evaluate the different models, a 4 folds-cross validation procedure was made, and the average of the performance indicators used were computed. By varying the size of the training data set, conclusions on the efficiency of the models in different conditions were presented. Three different ways for the random selection of points were chosen:

\begin{itemize}
\item The first consists in randomly choosing a proportion of points regardless of their location in space or when they were collected: this corresponds to the reconstruction of data between sampled places.
\item The second, more realistic, consists in choosing small paths of different lengths while keeping the same percentage of data in order to reproduce a real data collection from a mobile sensor: this corresponds to the extrapolation of the data to places close to the sampling places.
\item The last one, using only the data resulting from the trajectory of specific trams. This corresponds to extrapolation for places ``far'' from sampling points, what will often be encountered in practice.
\end{itemize}

\subsubsection{Performance Indicator}

The three approaches were compared to one deterministic interpolation technique here considered as the reference (IDW), in the three scenarios. The evaluation of the result of each of them used the following three performance indicators:

\begin{equation}
\text{BIAS} = \frac{1}{n}\sum_{i=1}^{n}(Z^*_i-Z_i)
\end{equation}

\begin{equation}
\text{RMSE} = \sqrt{\frac{\sum_{i=1}^{n}(Z^*_i-Z_i)^2}{n}}
\end{equation}

\begin{equation}
\text{CORR} = \frac{\sum_{i=1}^{n}(Z^*_i-\bar{Z^*})(Z_i-\bar{Z})}{\sqrt{\sum_{i=1}^{n}(Z^*_i-\bar{Z^*})^2\sum_{i=1}^{n}(Z_i-\bar{Z})^2}}
\end{equation}

\subsection{Data} \label{sec:data}

The OpenSense project \cite{aberer2010opensense}, is a Swiss project aiming to integrate air quality measurements from heterogeneous mobile and crowd sensed data sources in order to understand the health impacts of air pollution exposure, and to provide high-resolution urban air quality maps. It deployed several mobile air quality sensors on the trams' roofs in the Swiss city of Zurich and Lausanne's buses, collecting the measurement of ozone concentrations and counting Ultra Fine Particles (UFP). The data collection methodology can be found in these studies \cite{li2012sensing,hasenfratz2014pushing}. Even if these data show inconsistencies, especially the sampling only on static trajectories of the city, they remain nonetheless very valuable for the application and the evaluation of new approaches to model the spatio-temporal variability of pollution in the urban environment.

\begin{figure}[H]
\centering
\includegraphics[width=0.9\textwidth]{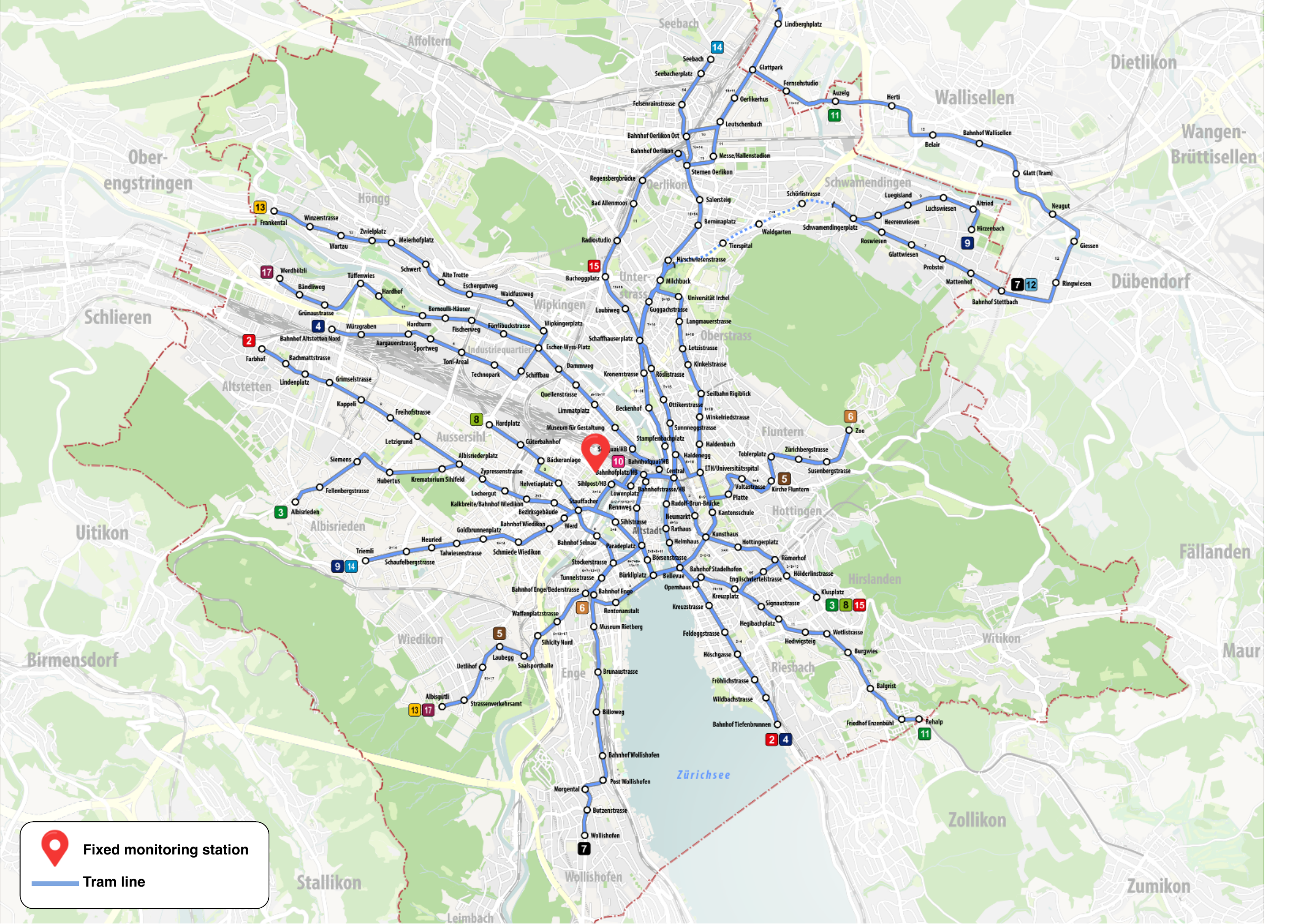}
\caption{Location of the fixed station and the tram path in the city of Zurich.}
\label{figure1}
\end{figure}

In this paper, the study was carried out using measured ozone concentration provided by the mobile sensors deployed on the top of the Zurich trams. The trajectory of the trams can be seen on Figure~\ref{figure1}.
Since the objective is to predict the concentration on a very detailed temporal resolution, this paper restricts the data used to a single week. The week from 28th February to 5th March 2016 has therefore been selected for demonstrating the mapping methodology.

The OpenSense data provide the ozone concentration in parts per billion (ppb) in a given volume (volume of gaseous pollutant per $10^9$ volumes of ambient air). In order to convert it to $\mu$g/m$^3$ to match the unit of the data from the fixed monitoring station we applied the following formula: $\mu$g/m$^3$ = (ppb) $\cdot$ (12.187) $\cdot$ (M)/293 where M is the molecular weight of the ozone pollutant ($M(O_3)=48$). An atmospheric pressure of 1 atmosphere and a temperature of 20$^\circ$C is assumed.

Reference data for fixed stations has been obtained from www.ostluft.ch, the official air quality monitoring network in eastern Switzerland, which manages several fixed stations in the country. The data used here is the ozone concentration, available hourly averaged. Since it is needed at a high temporal resolution, a linear interpolation was performed. The hourly averages were interpolated at each timestep when a measurement from the mobile sensors was collected.

The data provided by OpenSense are raw and not calibrated and a first analysis shows that the sensors measurement differ significantly from each other even when they are close to each other. To reduce the bias and errors, a linear transformation using the data from the fixed monitoring station considered as reference was applied. The calibration was carried out separately for each sensor in order to achieve the best possible performance for the various sensors without changing their respective correlation.

Precisely, the following model was used for the data:
\begin{equation}\label{eq:1}
Z_i=a_i+h_iY + \epsilon
\end{equation}
where $Z_i$ is the measurement of sensor $i$ at time $t$ and $Y$ is the measurement of the fixed station. $a$ and $h$ are respectively the additive bias and the multiplicative bias and $\epsilon$ errors with zero mean. The identification of $a$ and $h$ for each of the sensors was done using the ordinary least squares method.

\section{Kriging Theory and Inverse Distance Weighting} \label{sec:theory}

There are two ways of incorporating time into spatial geostatistics. The first one in the form of cokriging, the second, more natural, by considering time as a separate dimension, which will be the case in this study.

What has been considered here as support, is a unique sample measured in a volume of air.

Given a support $D$ in $\mathbb{R}^n$ and a probability space $(\Omega, \mathcal{A}, \mathbb{P})$, a random function is a function of two variables $Z(x,w)$ such that for each $x$ in $D$ the section $Z(x,.)$ is a random variable on $(\Omega, \mathcal{A}, \mathbb{P})$.

In this case $D=\mathbb{R}^2\times\mathbb{R}_+$ where $\mathbb{R}^2$ represents space and $\mathbb{R}_+$ time, the random function is simply denoted by $Z(x,t)$ and a realisation of this random function is represented by $z(x,t)$ where $x\in \mathbb{R}^2$ and $t\in \mathbb{R}_+$.

\subsection{Simple Kriging with a Varying Mean}

The application of simple kriging requires two hypotheses: the second order stationarity of the random field, and the knowledge of the mean $m$ over the whole domain $D$.
Assuming that the data collected directly by the mobile sensors come from a stationary random field of order two is a strong hypothesis which is not realistic. In this paper, the data given by the fixed monitoring station $F(t)$ is supposed to be the overall mean.
Subtracting the value of the fixed station from the data provided by the mobile sensors: $Z(x,t)=Y(x,t)-F(t)$ is assumed to be stationary of order two with zero mean.

The simple kriging estimator is:
\begin{equation} \label{eq:2}
Z^*(x,t) = \mu + \sum_{i=1}^{n} \lambda_i (Z(x_i,t_i) - \mu) = \sum_{i=1}^{n} \lambda_i Z(x_i,t_i)
\end{equation}

To produce the best linear estimator, we must ensure that the estimation variance is minimal and that the estimator is unbiased.
The unbiased condition is automatically verified, and does not imply any additional constraint because:
\begin{equation} \label{eq:3}
\mathbb{E}[Z^*(x,t)-Z(x,t)] = \sum_{i=1}^{n}\lambda_i\mathbb{E}Z(x_i,t_i) = 0
\end{equation}

This leads to the simple kriging equations:
\begin{equation} \label{eq:4}
\sum_{j=1}^{n}\lambda_j\gamma(x_i-x_j,t_i-t_j) = \gamma(x_i-x,t_i-t) \qquad i=1,\ldots,n
\end{equation}

The resolution of equations \eqref{eq:4} gives the different $\lambda$ in the linear combination \eqref{eq:2}.

\subsection{Ordinary Kriging}

The application of ordinary kriging makes a little less restrictive assumptions, namely a constant but unknown mean.
The linear estimator of ordinary kriging is written this way:
\begin{equation} \label{eq:5}
Z^*(x,t) = \sum_{i=1}^{n} \lambda_i Z(x_i,t_i)
\end{equation}

To ensure the unbiased condition:
\begin{equation} \label{eq:6}
\mathbb{E}[Z^*(x,t)] = \mathbb{E}\left[\sum_{i=1}^{n} \lambda_i Z(x_i,t_i)\right] = m\sum_{i=1}^{n} \lambda_i
\end{equation}
\begin{equation} \label{eq:7}
\sum_{i=1}^{n} \lambda_i = 1
\end{equation}

The objective is to minimize $\sigma^2$ under the unbiased condition \eqref{eq:7} using the Lagrangian multiplier $\mu$.
The weights that minimize $\sigma^2$ are the solution of:
\begin{align}
\label{eq:8}
\sum_{j=1}^{n}\lambda_j\gamma(x_i-x_j,t_i-t_j) + \mu &= \gamma(x_i-x,t_i-t) \qquad i=1,\ldots,n \\
\sum_{i=1}^{n} \lambda_i &= 1 \nonumber
\end{align}

The equation system \eqref{eq:8} is called the ordinary kriging system, solving it yields the weights $\lambda_i$ for the linear estimator \eqref{eq:5}.

\subsection{Kriging with External Drift}

Kriging with external drift or regression kriging assumes that $Z(x,t)$ can be broken down into two parts, one deterministic $\mu(x,t)$ and the other stochastic $Y(x,t)$:
\begin{equation} \label{eq:9}
Z(x,t) = \mu(x,t) + Y(x,t)
\end{equation}
with $Y$ being stationary intrinsic with zero mean.

$f_0, f_1, \ldots, f_L$ are deterministic functions with $f: D \longrightarrow \mathbb{R}$
and $\mu(x,t)$ is a linear combination of these functions evaluated at $(x,t)$:
\begin{equation} \label{eq:10}
\mu(x,t) = \sum_{l=0}^{L}a_l f_l(x,t)
\end{equation}
with $f_0(x,t)=1$
\begin{equation}\label{eq:11}
Z(x_i,t_i) = \mu(x_i,t_i) + Y(x_i,t_i) = \sum_{l=0}^{L}a_l f_l(x_i,t_i) + Y(x_i,t_i)
\end{equation}

The different functions $f_l(x,t)$ represent the covariates ``external drifts'' used to estimate the underlying mean, in this study, only one function $f_1(x,t)=F(t)$ which stands for the fixed stations data was used.

The linear kriging with external drifts estimator is therefore written:
\begin{equation}\label{eq:12}
Z^*(x,t) = \sum^{n}_{i=1}w_i Z(x_i,t_i) = \sum^{n}_{i=1}w_i\left(\sum_{l=0}^{1}a_l f_l(x_i,t_i) + Y(x_i,t_i)\right)
\end{equation}

The unbiased condition is satisfied if and only if:
\begin{equation} \label{eq:13}
\sum_{i=1}^{n}w_i f_l(x_i,t_i) = f_l(x,t) \qquad l=0,1
\end{equation}

Coupled with the minimum variance condition, it gives the kriging system \eqref{eq:14}:
\begin{align}
\label{eq:14}
\sum_{j=1}^{n}\lambda_j\gamma(x_i-x_j,t_i-t_j) + \sum_{l=0}^{1}a_l f_l(x_i,t_i) &= \gamma(x_i-x,t_i-t) \qquad i=1,\ldots,n \\
\sum_{i=1}^{n}w_i f_l(x_i,t_i) &= f_l(x,t) \qquad l=0,1 \nonumber
\end{align}

\subsection{Spatio-Temporal Inverse Distance Weighting}

Inverse Distance Weighting is a type of deterministic method, which assigns values to non-sampled points using a linear combination of values from sampled points weighted by the inverse distance.

The general formula for the IDW is given by equation \eqref{eq:15}:
\begin{equation} \label{eq:15}
Z^*(x,t) = \sum_{i=1}^n \lambda_i z(x_i,t_i)
\end{equation}
with:
\begin{equation} \label{eq:16}
\lambda_i = \frac{1/d_i^p}{\sum_{i=1}^{n} 1/d_i^p}
\end{equation}

$d_i$ represents the distance between $Z^*(x,t)$ and $z(x_i,t_i)$.
The weights decrease as the distance increases, especially as the power value $p$ rises. Like the previous methods, points in the neighbourhood have a heavier weight and have more influence on the prediction, resulting in a local spatio-temporal interpolation.
In this study, this definition of a spatio-temporal distance was chosen:
\begin{equation} \label{eq:17}
d_{i} = \sqrt{(x_{i}-x)^2 + (y_{i}-y)^2 + C \cdot (t_{i}-t)^2}
\end{equation}

The parameter $p$ was fixed at 2, while $C$ was obtained by cross-validation.

Finally, it is important to note that, while any covariance function can be written in the form of a variogram using $\gamma(h) = C(0) - C(h)$, the opposite is not generally true. The passage from variogram to covariance is only possible under the assumptions of second order stationarity.

This paper only uses the variogram and not the covariance function, making less strict assumptions.

\section{Results}

In this chapter, different results from the application of the methodology on the dataset will be presented. Starting with the variographic study, by showing the two different directional variograms, as well as the experimental variograms and their respective theoretical variograms considered for the different models.
Then, a prediction with the three modelling was carried out for the day of 2016.03.04 from 5 am to 10 pm. These data were chosen because they represent typical daily ozone variation with a peak around 2 pm.
The result of the cross validation procedure in each of the scenarios is shown for the three performance indicators considered for the three spatio-temporal geostatistical modelling as well as the IDW method.
Last, the prediction of ozone concentration during a day from all the data available using the kriging with external drift model is displayed.

\begin{figure}[H]
\centering
\begin{tabular}{cc}
\includegraphics[width=6.5cm]{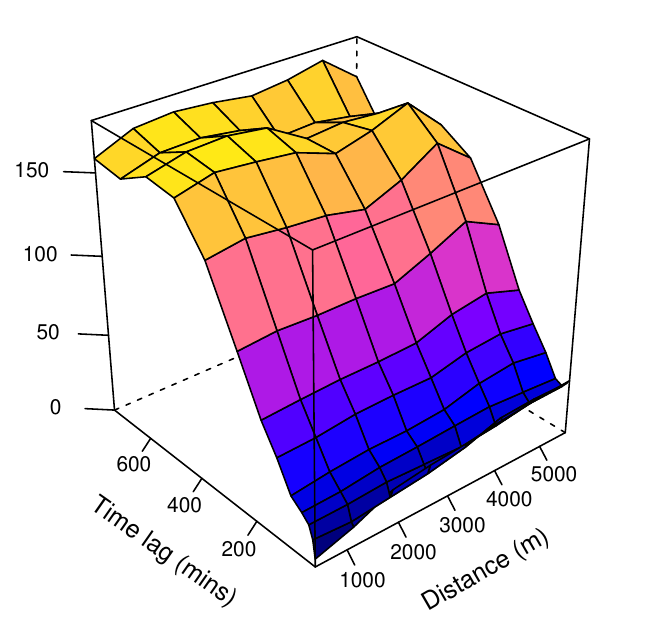} & \includegraphics[width=6.5cm]{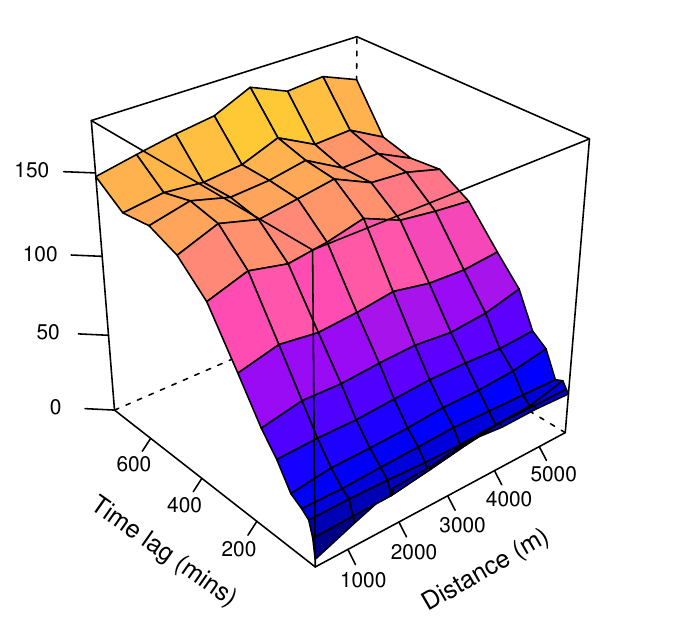} \\
(a) East-West axis & (b) North-South axis
\end{tabular}
\caption{Directional spatio-temporal empirical variograms.}
\label{directional_vario}
\end{figure}

\subsubsection{Spatio-Temporal Variance}

\begin{figure}[H]
\centering
\begin{tabular}{cc}
\includegraphics[width=6.5cm]{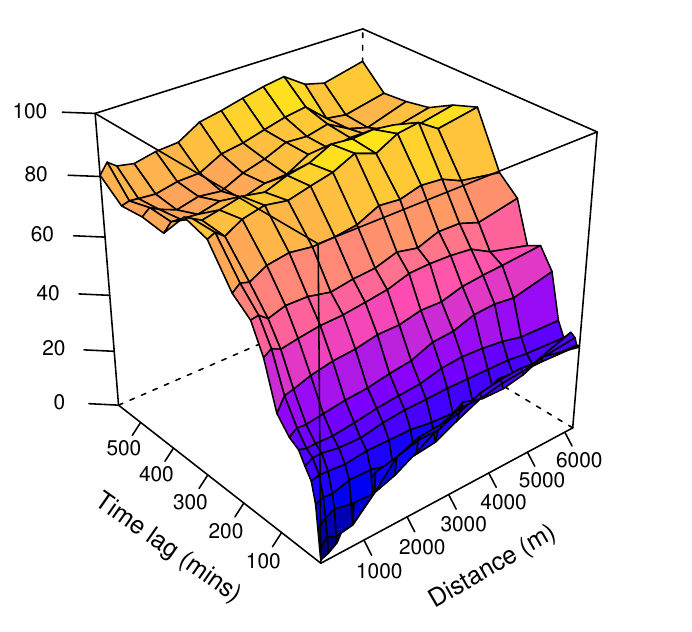} & \includegraphics[width=6.5cm]{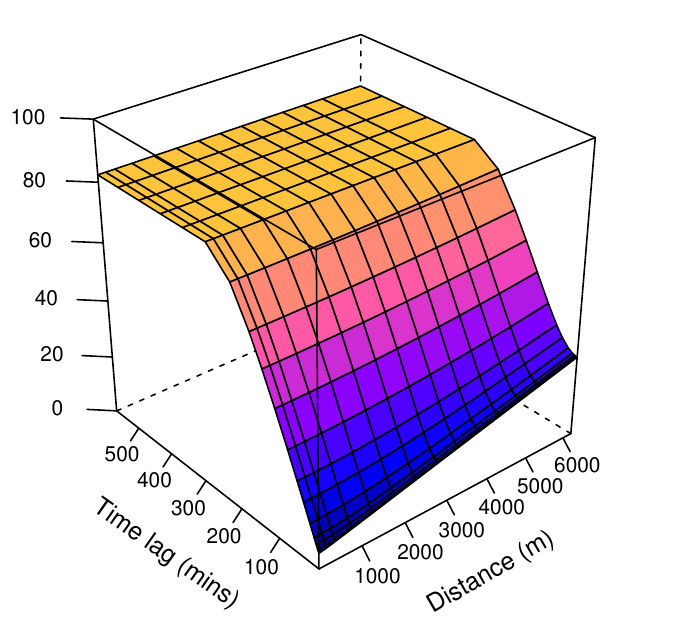} \\
(a) Empirical variogram & (b) Theoretical model
\end{tabular}
\caption{Spatio-temporal variograms associated with the simple kriging model.}
\label{vario_SK}
\end{figure}

\begin{figure}[H]
\centering
\begin{tabular}{cc}
\includegraphics[width=6.5cm]{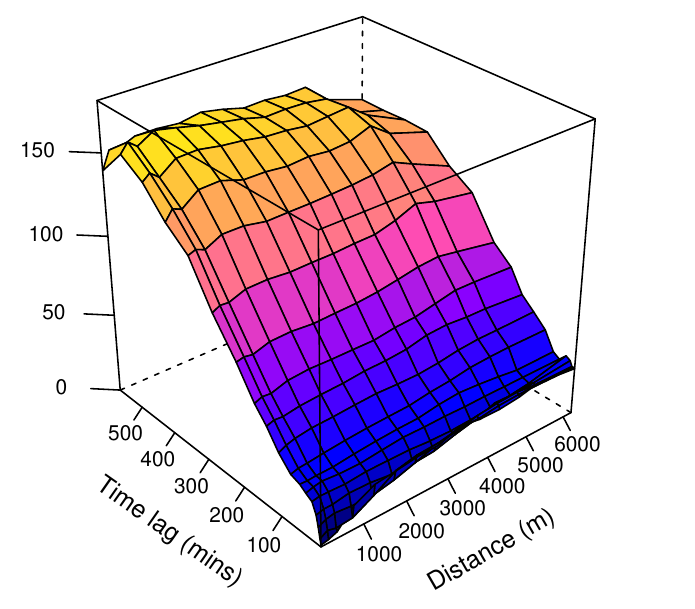} & \includegraphics[width=6.5cm]{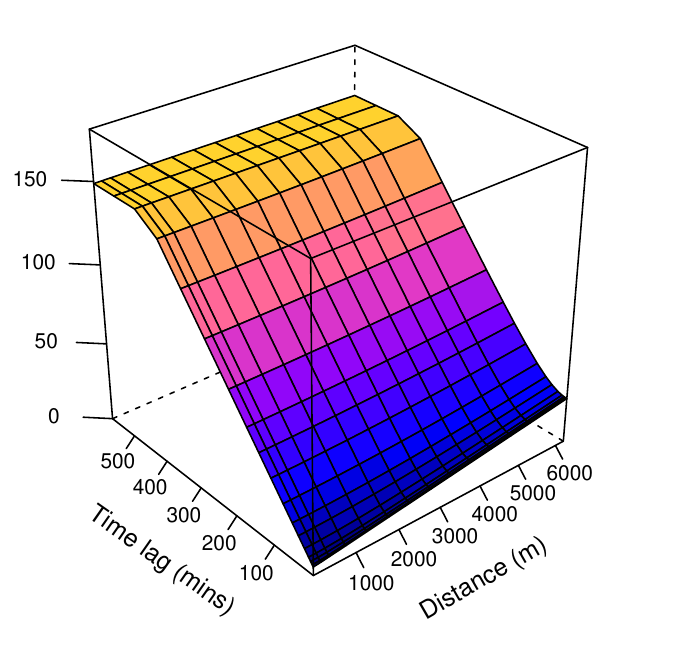} \\
(a) Empirical variogram & (b) Theoretical model
\end{tabular}
\caption{Spatio-temporal variograms associated with the ordinary kriging model.}
\label{vario_OK}
\end{figure}

\begin{figure}[H]
\centering
\begin{tabular}{cc}
\includegraphics[width=6.5cm]{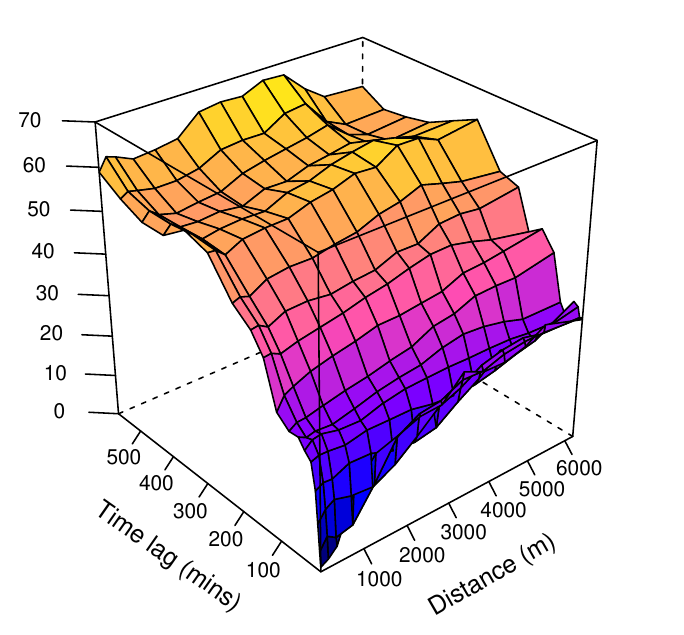} & \includegraphics[width=6.5cm]{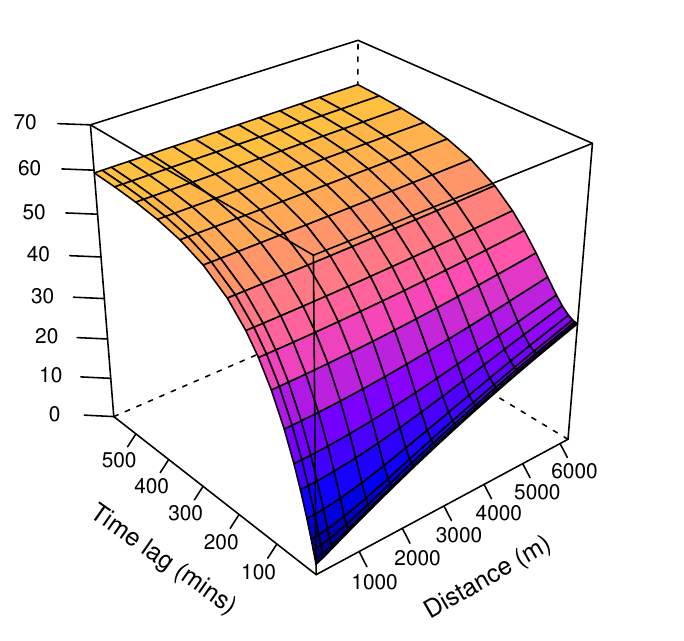} \\
(a) Empirical variogram & (b) Theoretical model
\end{tabular}
\caption{Spatio-temporal variograms associated with the Kriging with external drift model.}
\label{vario_UK}
\end{figure}

The study of the spatio-temporal variance of the data shows a clear difference between the spatial and temporal variability. The different variograms show that, in average, there is a greater difference between two measurements sampled a few hours apart at the same place, than two measurements sampled at the same time anywhere in space (on the scale of a city) which justifies the traditional approach using the fixed stations for monitoring air quality. Mobile sensors, in addition to be able to capture temporal variance, can also capture spatial variance. Since we do not sacrifice spatial variance by using them, we can only improve the explained global variance.
The three variograms (raw data in Figure~\ref{vario_OK}, residuals in Figure~\ref{vario_SK} and estimated residuals in Figure~\ref{vario_UK}) show exactly the same purely spatial variability. This is because for residual variograms, we have subtracted only temporal component provided by the fixed monitoring station, leaving the spatial variability unchanged.

The three computed empirical variograms show small nugget effects, but there is no data at the same time and at the same place simultaneously as none of the trams meet. Moreover, the proximate collected data points necessarily come from the same sensor and these measurements are not independent conditionally to the ozone concentration. This is why these variograms show a little variability in the origin, which does not necessarily reflect the real variability of the studied phenomenon.

\subsubsection{Modelling}

A metric theoretical spatio-temporal variogram assumes identical spatial and temporal covariance functions taking into account spatio-temporal anisotropy:
\begin{equation}
\gamma(h, u) = \gamma_{\text{joint}}\left(\sqrt{h^2 + (K \cdot u)^2}\right)
\end{equation}
where $\gamma_{\text{joint}}$ is any known variogram that may include a nugget effect and $K$ is a spatio-temporal anisotropy parameter defined as the number of space units equivalent to one time unit. The different joint models and their respective parameters can be found in Table~\ref{tab2} for the three methods.

\begin{table}[H]
\centering
\caption{Different joint models and their respective parameters.}
\label{tab2}
\begin{tabular}{ccccccc}
\toprule
\textbf{Method} & \textbf{S-P model} & \textbf{K} & \textbf{Joint model} & \textbf{Sill} & \textbf{Nugget} & \textbf{Range} \\
\midrule
Simple kriging & Metric & 105.16 & Spheric & 82.30 & 5.00 & 30415.43 \\
Ordinary kriging & Metric & 91.18 & Linear & 148.8 & 5.00 & 38073.4 \\
Kriging with external drift & Metric & 83.03 & Exponential & 59.86 & 2.00 & 9872.405 \\
\bottomrule
\end{tabular}
\end{table}

As expected, the model associated with the ordinary kriging shows the highest range and sill, as opposed to the two other models, where the data from the fixed station partially explained the variance, resulting in lower range and sill.

\subsection{Spatio-Temporal Signals}

\begin{figure}[H]
\centering
\begin{tabular}{cc}
\includegraphics[width=6.5cm]{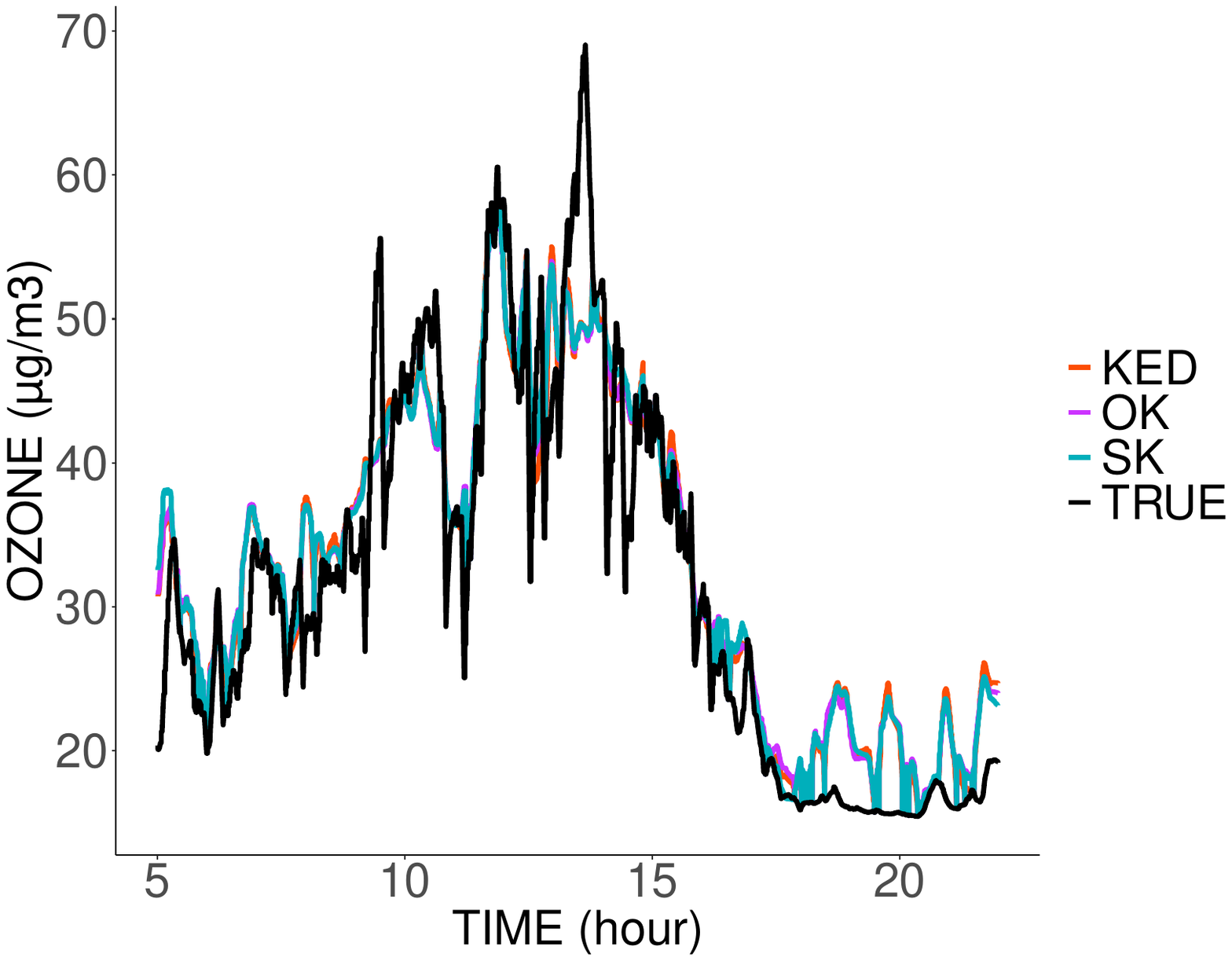} & \includegraphics[width=6.5cm]{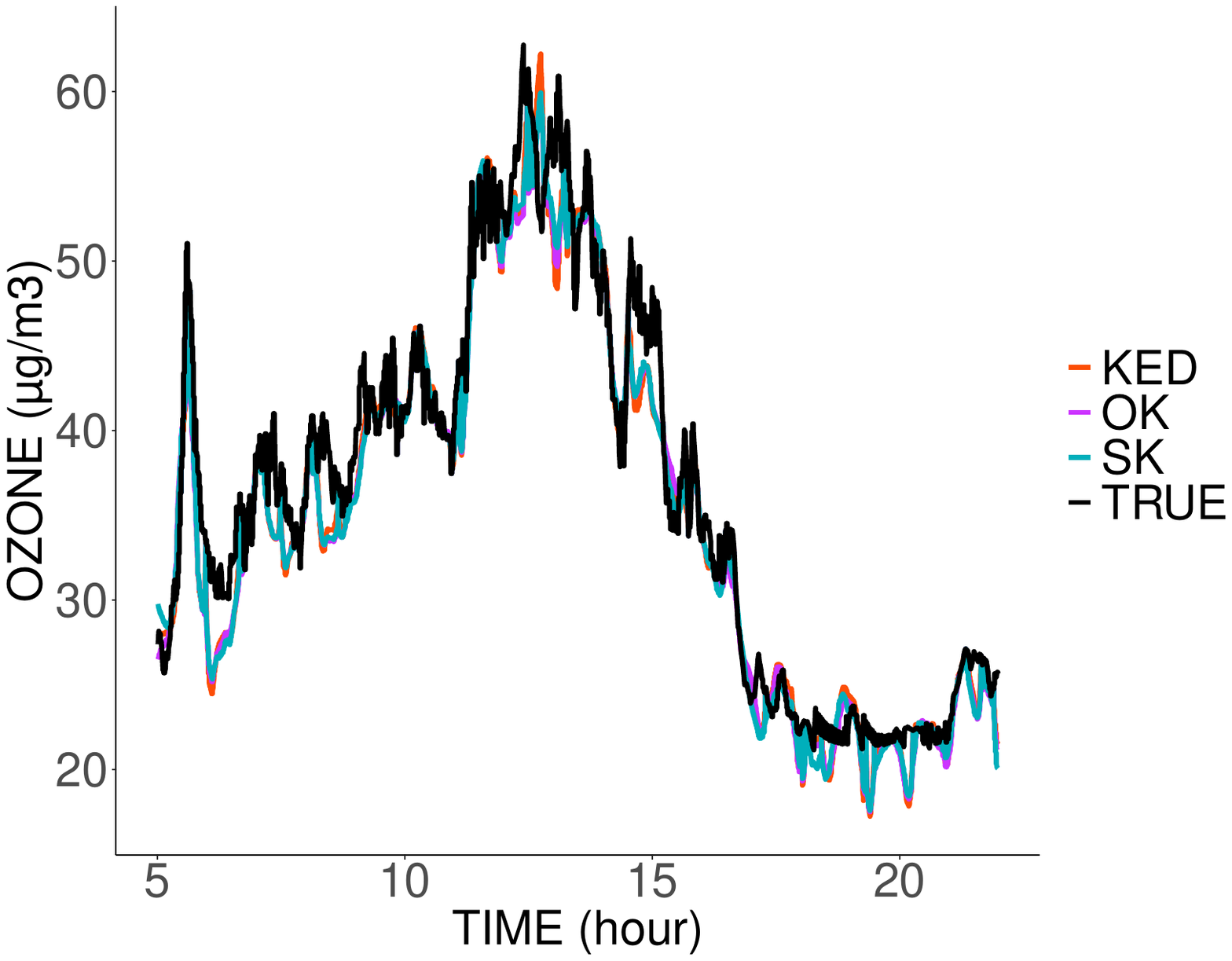} \\
(a) Line 4 & (b) Line 7 \\
\includegraphics[width=6.5cm]{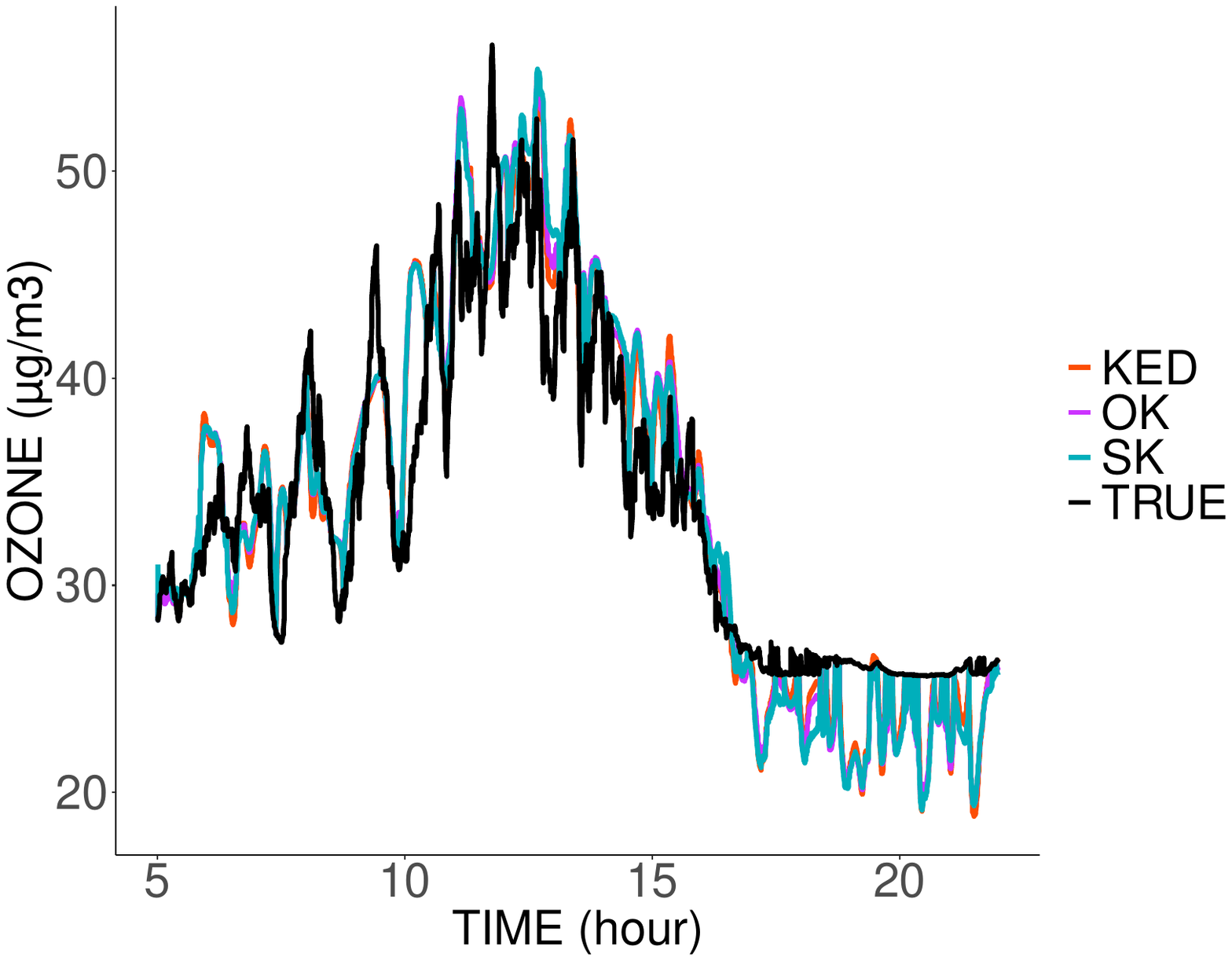} & \includegraphics[width=6.5cm]{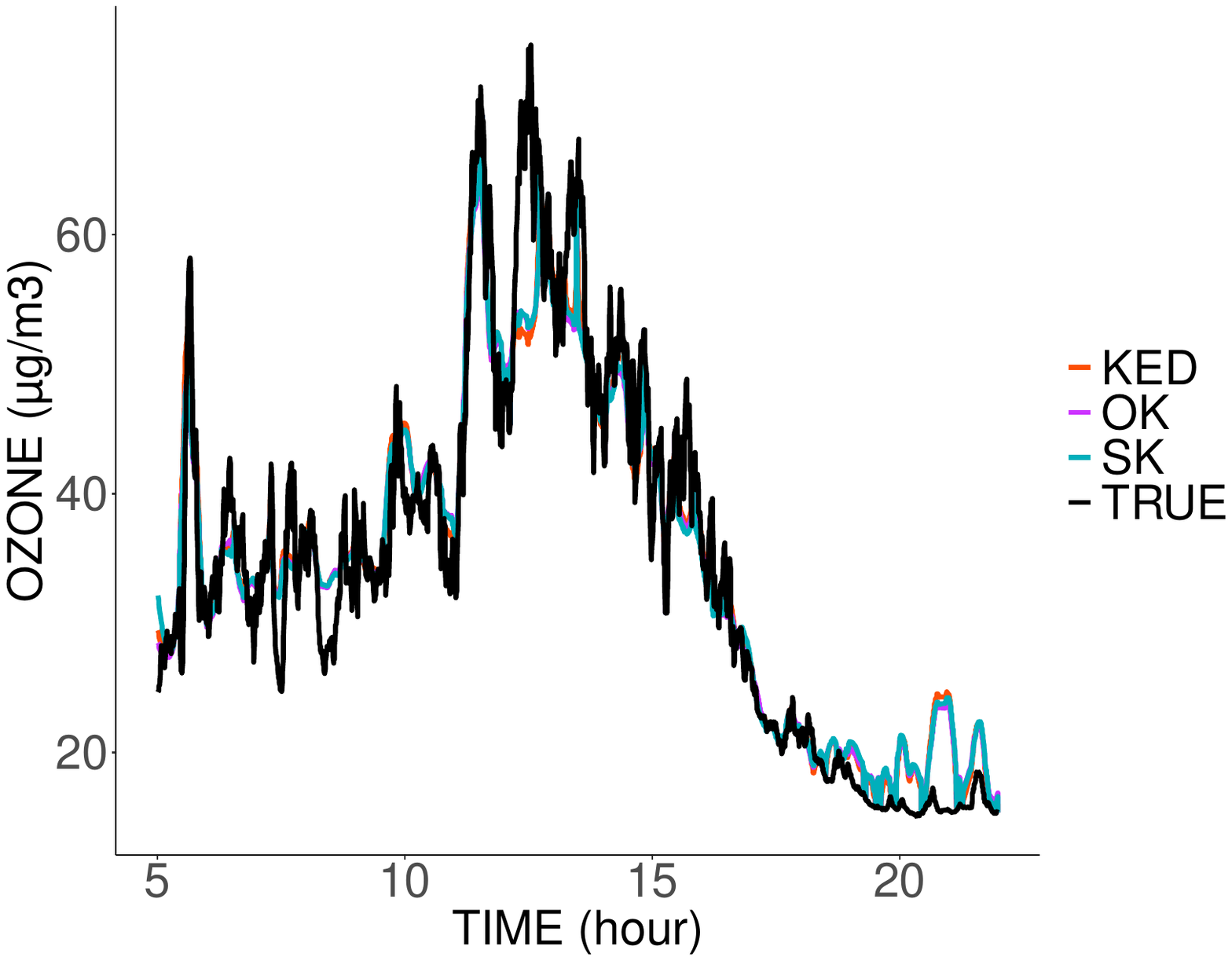} \\
(c) Line 8 & (d) Line 13 \\
\end{tabular}
\caption{Comparison between prediction and real value from four lines of tram in 04/03/2016 in the first scenario.}
\label{Scenario1}
\end{figure}

\begin{figure}[H]
\centering
\begin{tabular}{cc}
\includegraphics[width=6.5cm]{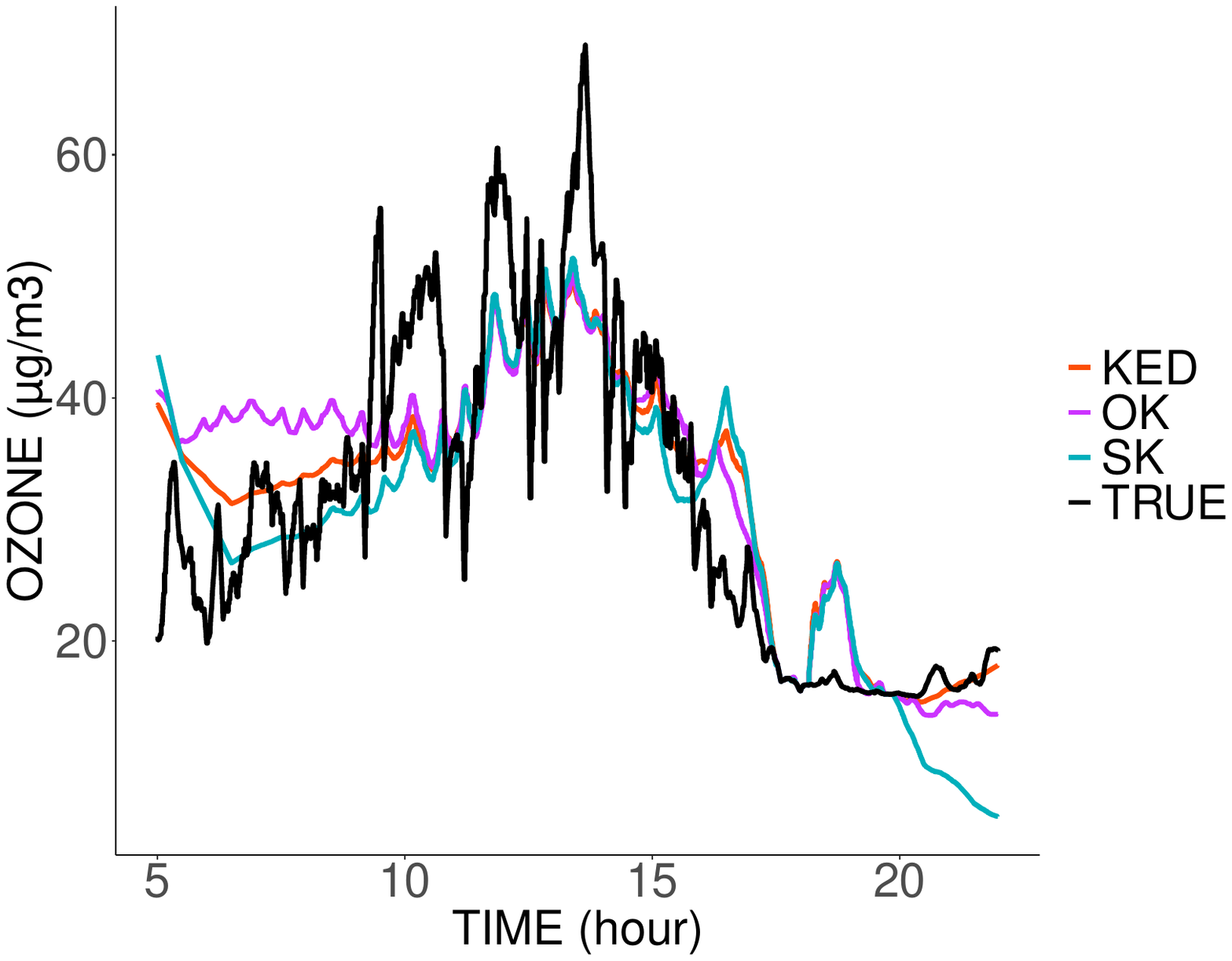} & \includegraphics[width=6.5cm]{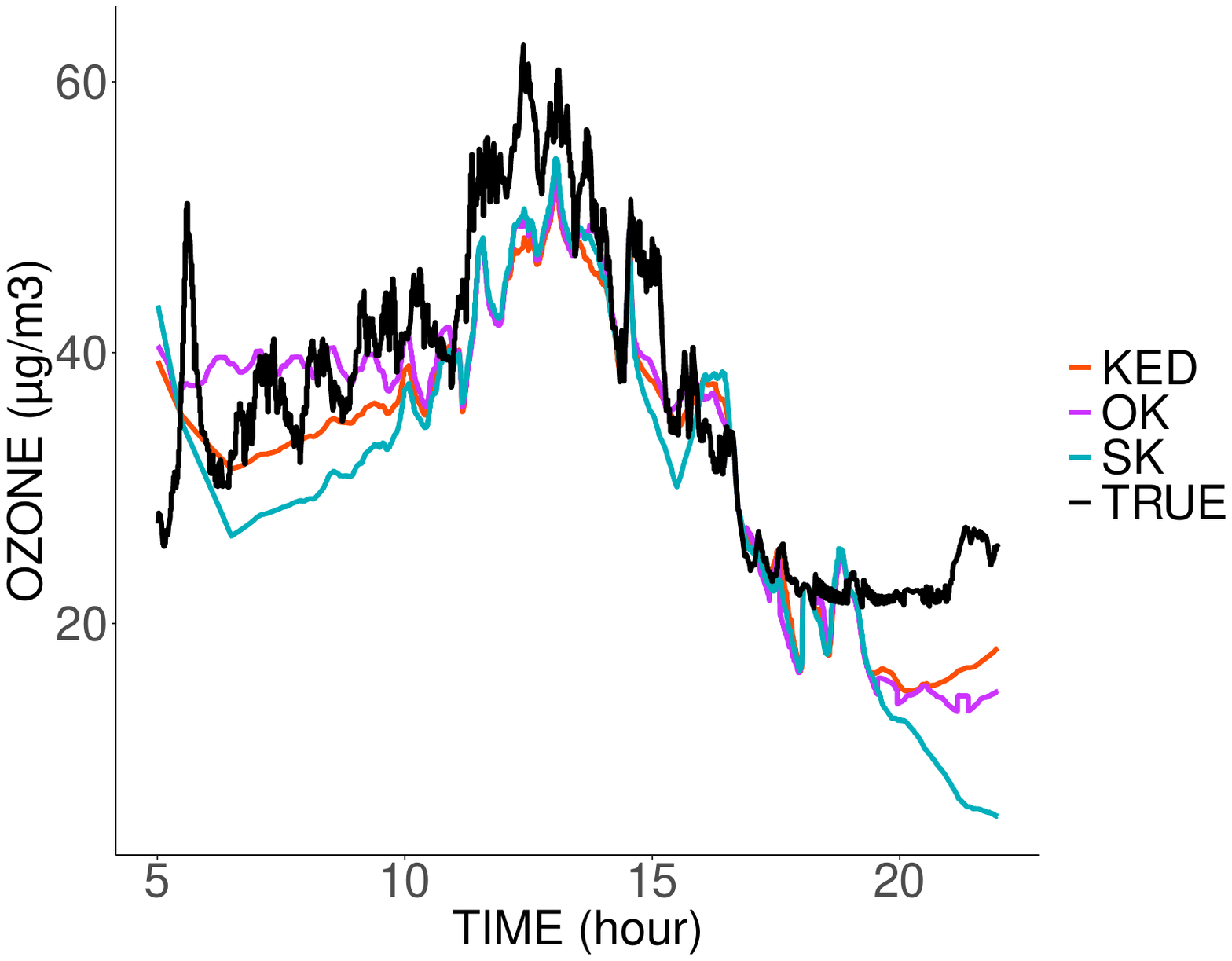} \\
(a) Line 4 & (b) Line 7 \\
\includegraphics[width=6.5cm]{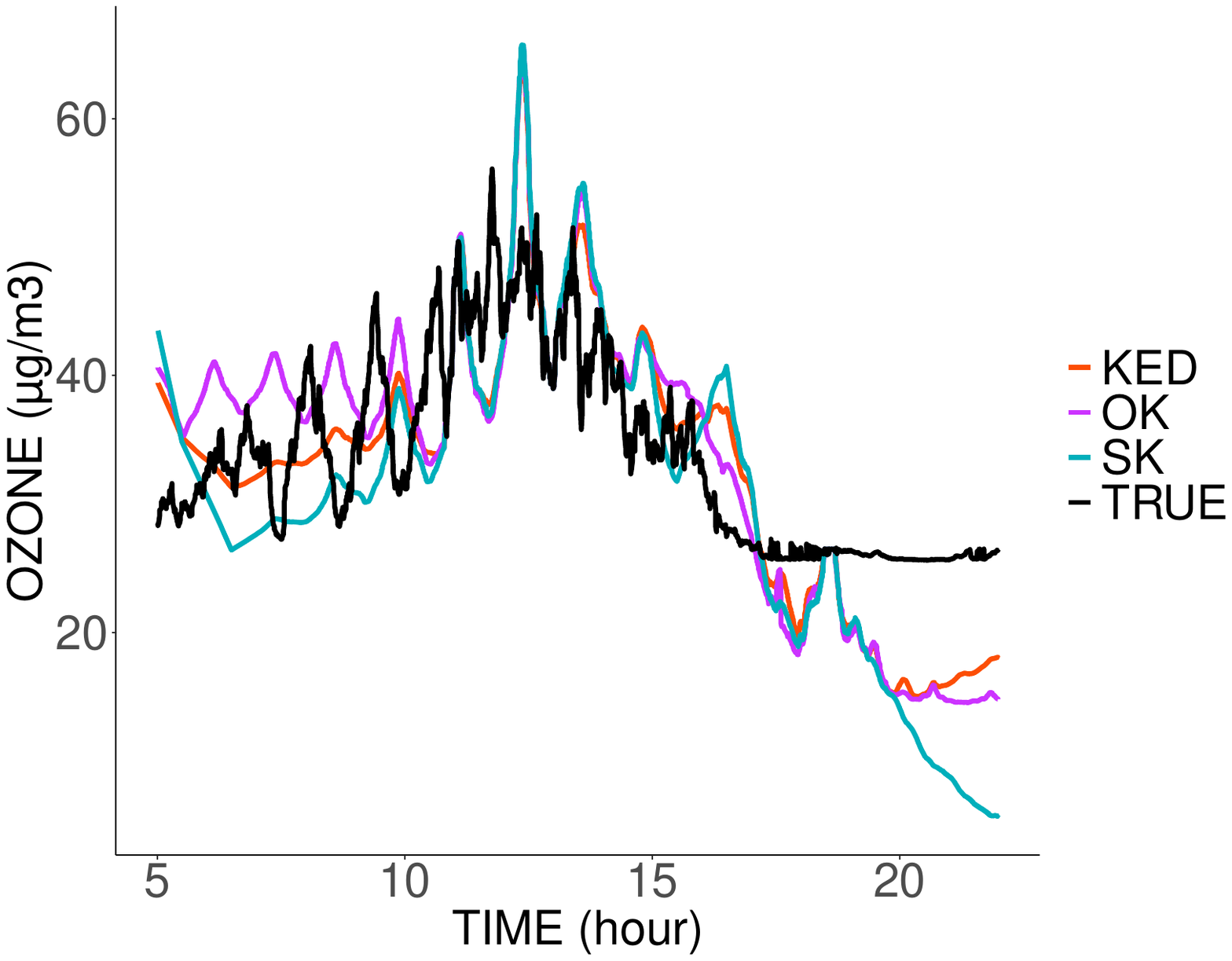} & \includegraphics[width=6.5cm]{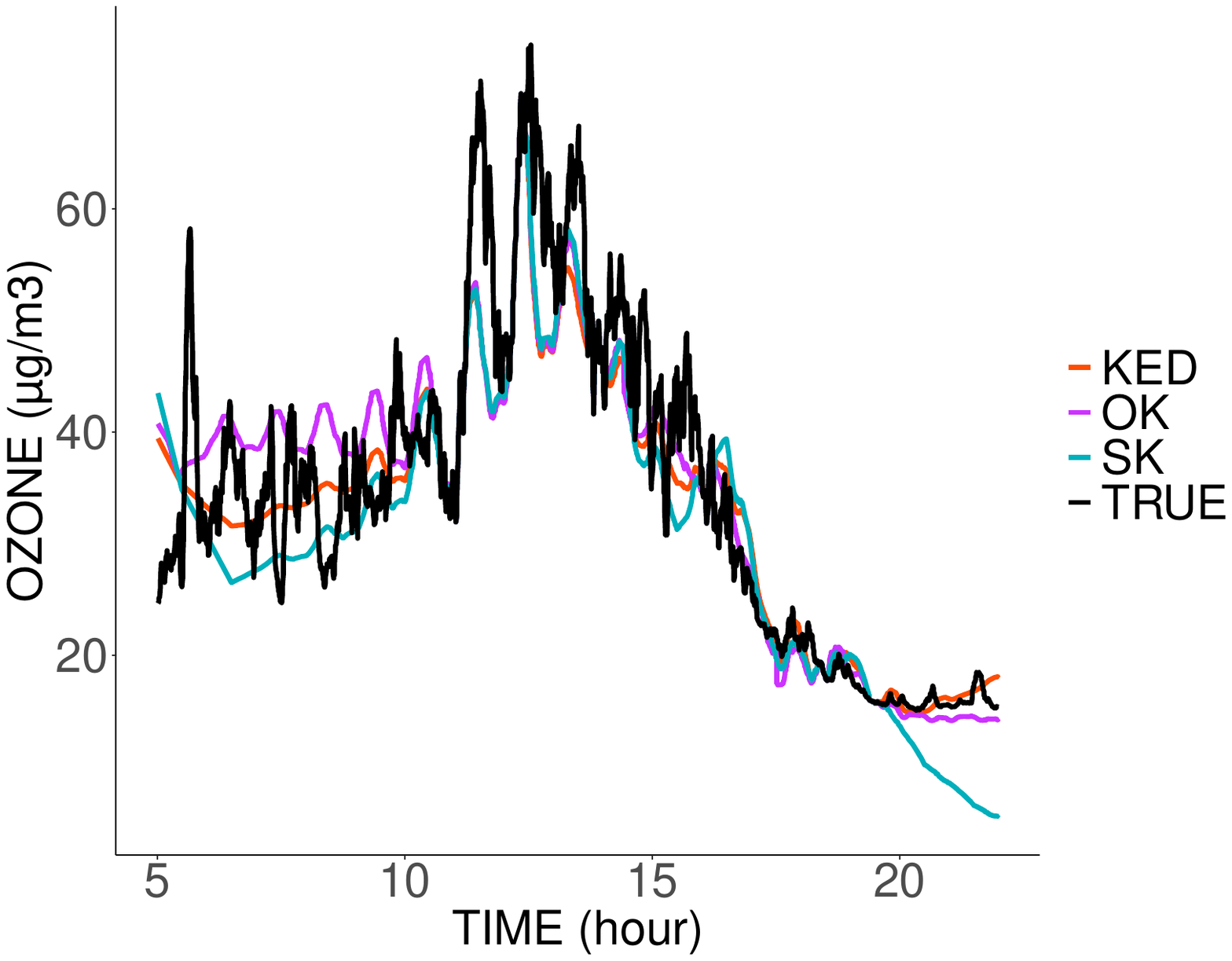} \\
(c) Line 8 & (d) Line 13 \\
\end{tabular}
\caption{Comparison between prediction and real value from four lines of tram in 04/03/2016 in the second scenario.}
\label{Scenario2}
\end{figure}

\begin{figure}[H]
\centering
\begin{tabular}{cc}
\includegraphics[width=6.5cm]{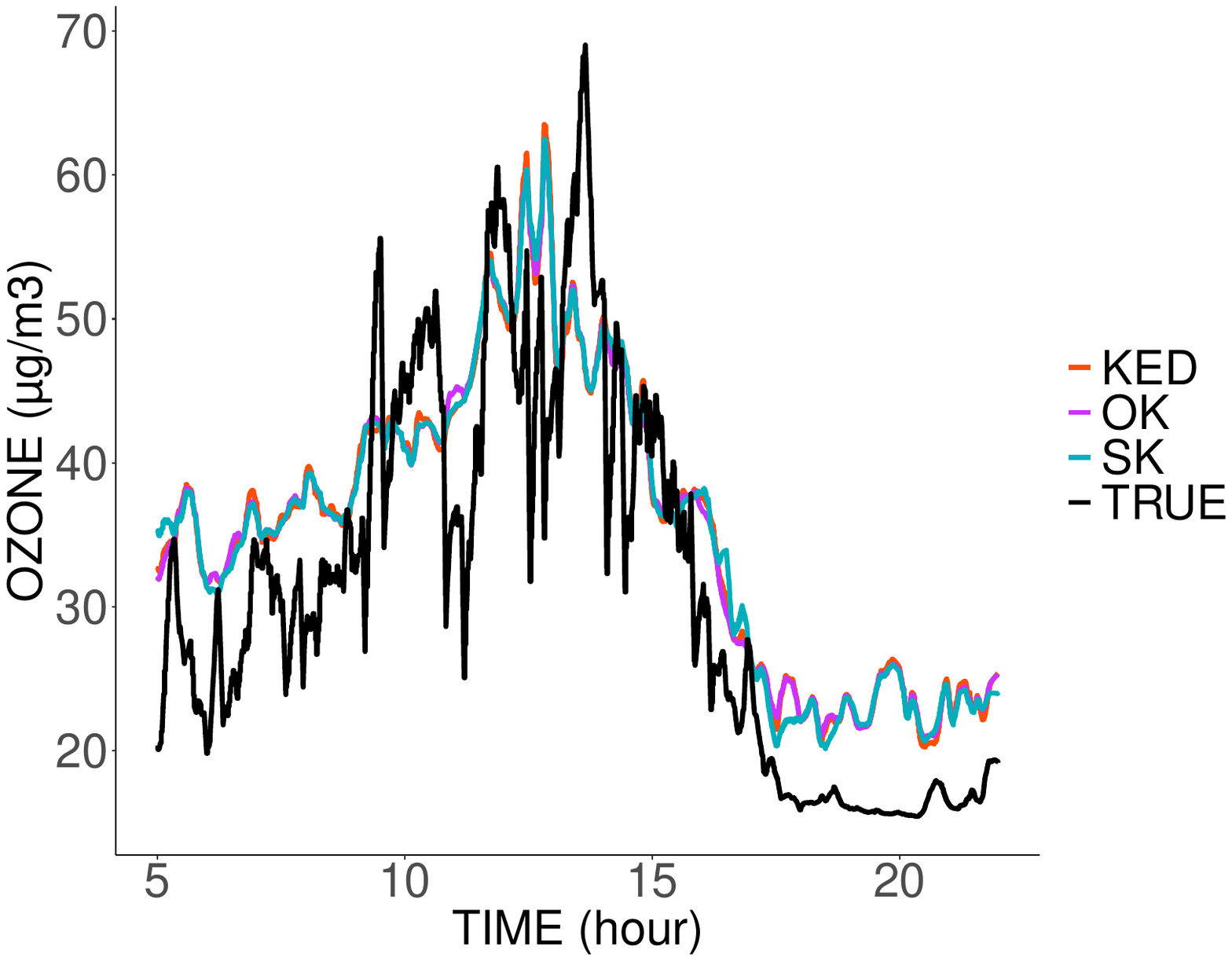} & \includegraphics[width=6.5cm]{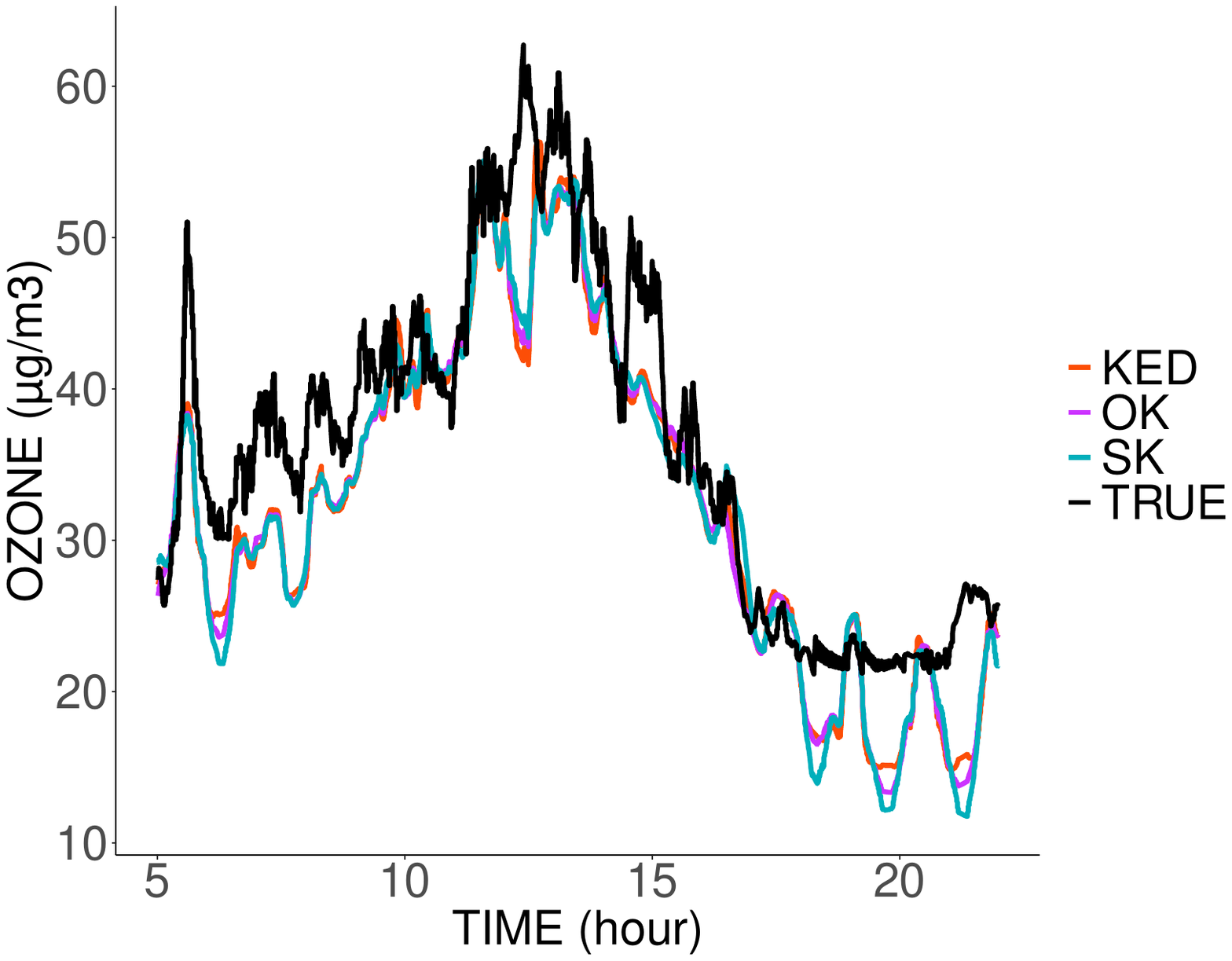} \\
(a) Line 4 & (b) Line 7 \\
\includegraphics[width=6.5cm]{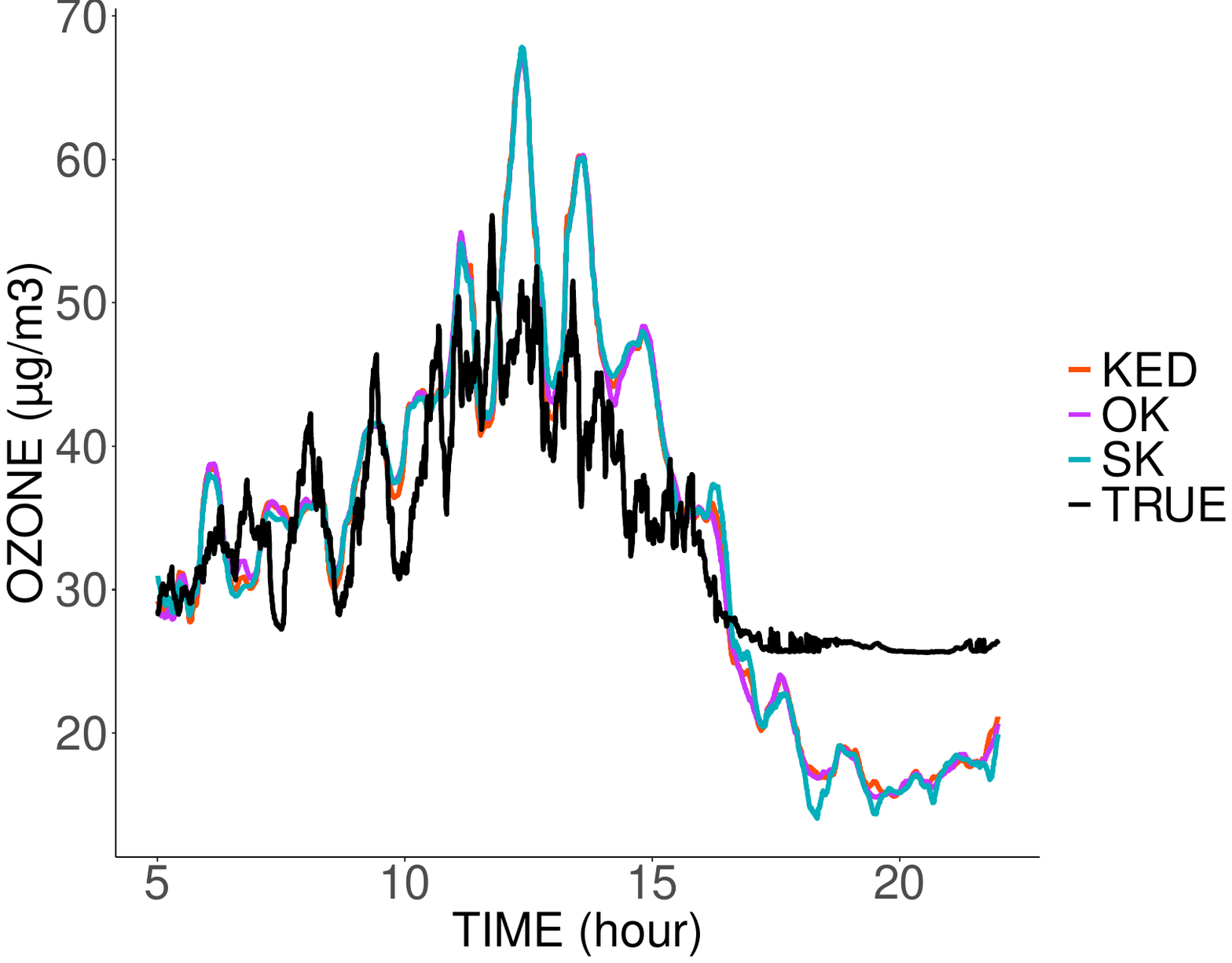} & \includegraphics[width=6.5cm]{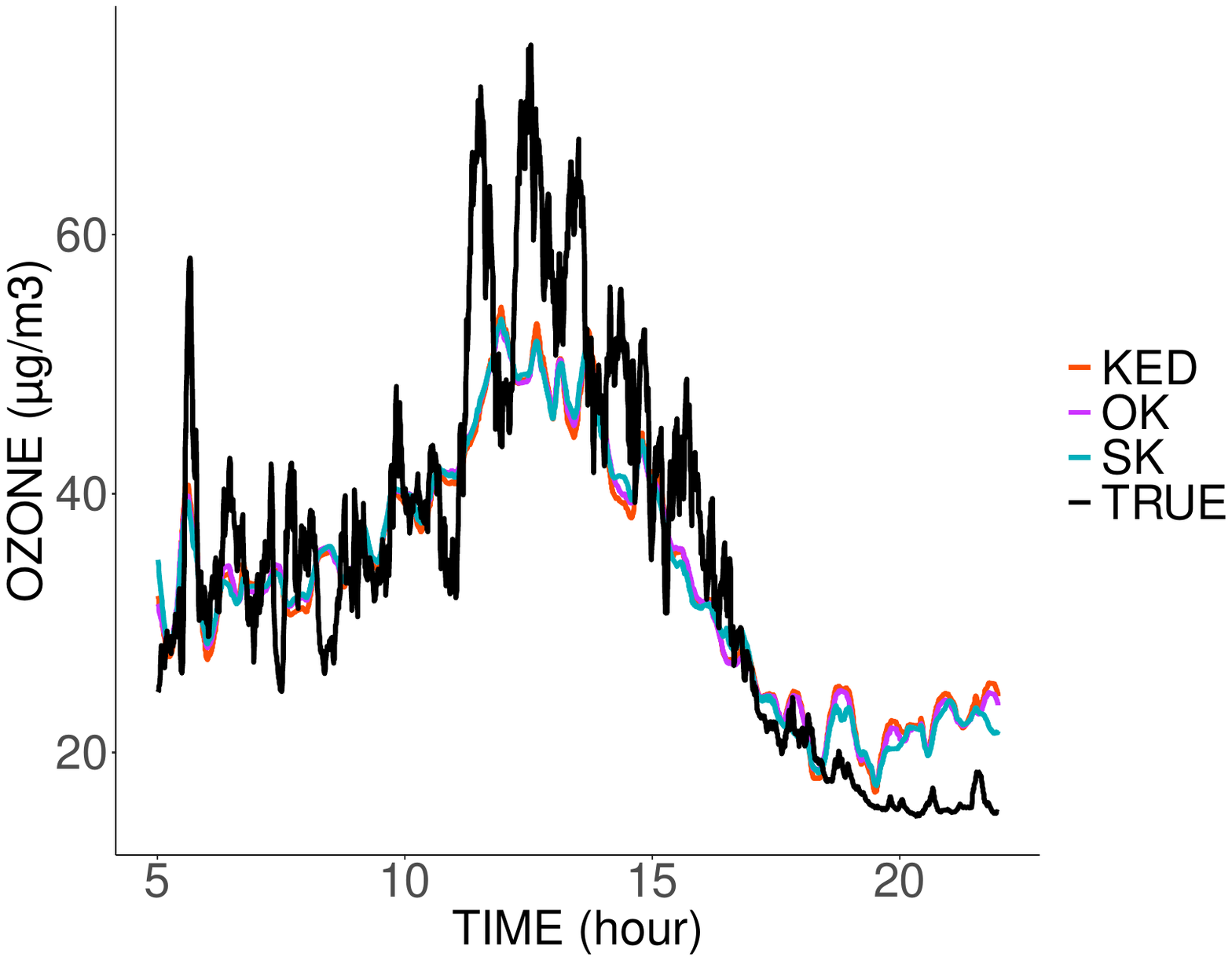} \\
(c) Line 8 & (d) Line 13 \\
\end{tabular}
\caption{Comparison between prediction and real value from four lines of tram in 04/03/2016 in the third scenario.}
\label{Scenario3}
\end{figure}

Figures \ref{Scenario1}, \ref{Scenario2} and \ref{Scenario3} show the prediction for the different tram lines trajectories. In the first, second and third scenarios respectively. Only four sensors were functional that day: the sensors on the lines 4, 7, 8 and 13.

The first thing to notice is the similarity of the predictions of the three methods. This is explained by the same spatial variability common to the 3 variograms. Moreover, this spatial variability is smaller than the temporal one, so the 3 estimators mainly use the data spatially close. As the spatial variability does not change from one model to another, we find a fairly similar prediction.
The three estimators do not interpolate the data at the sampled locations, they are therefore not exact estimators due to the nugget effect which represents measurement errors. The estimators therefore tend to filter the measurement errors.

In the third scenario (Figure~\ref{Scenario3}), in the absence of data coming from the predicted tram line, the estimators tends to imitate the values sampled in the nearest tram lines. Thus, the prediction on line 8 is similar to the values sampled on line 13, and vice versa.

The inadequacy of predictions at a given location comes from the lack of nearby data at that location, this is more visible in scenario 3.
The result is even worse at the end of the day. Indeed, in the absence of close data from the same tram, the predictions will be more influenced by the measurements taken at the same time by the other trams, but we notice a clear difference in the measurements taken at the end of the day.

\subsection{Performance Indicators Results}

\begin{figure}[H]
\centering
\begin{tabular}{ccc}
\includegraphics[width=4.2cm]{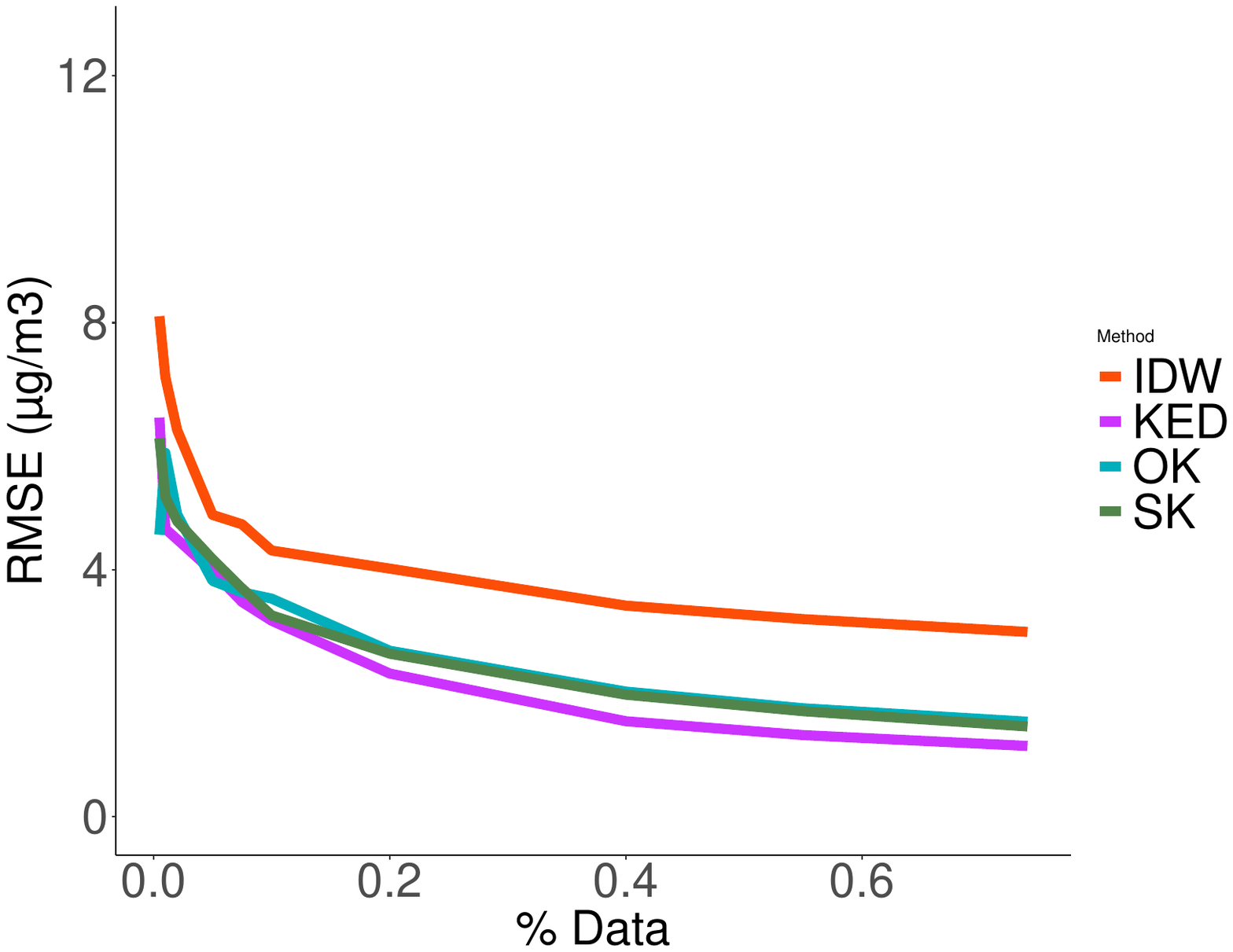} & \includegraphics[width=4.2cm]{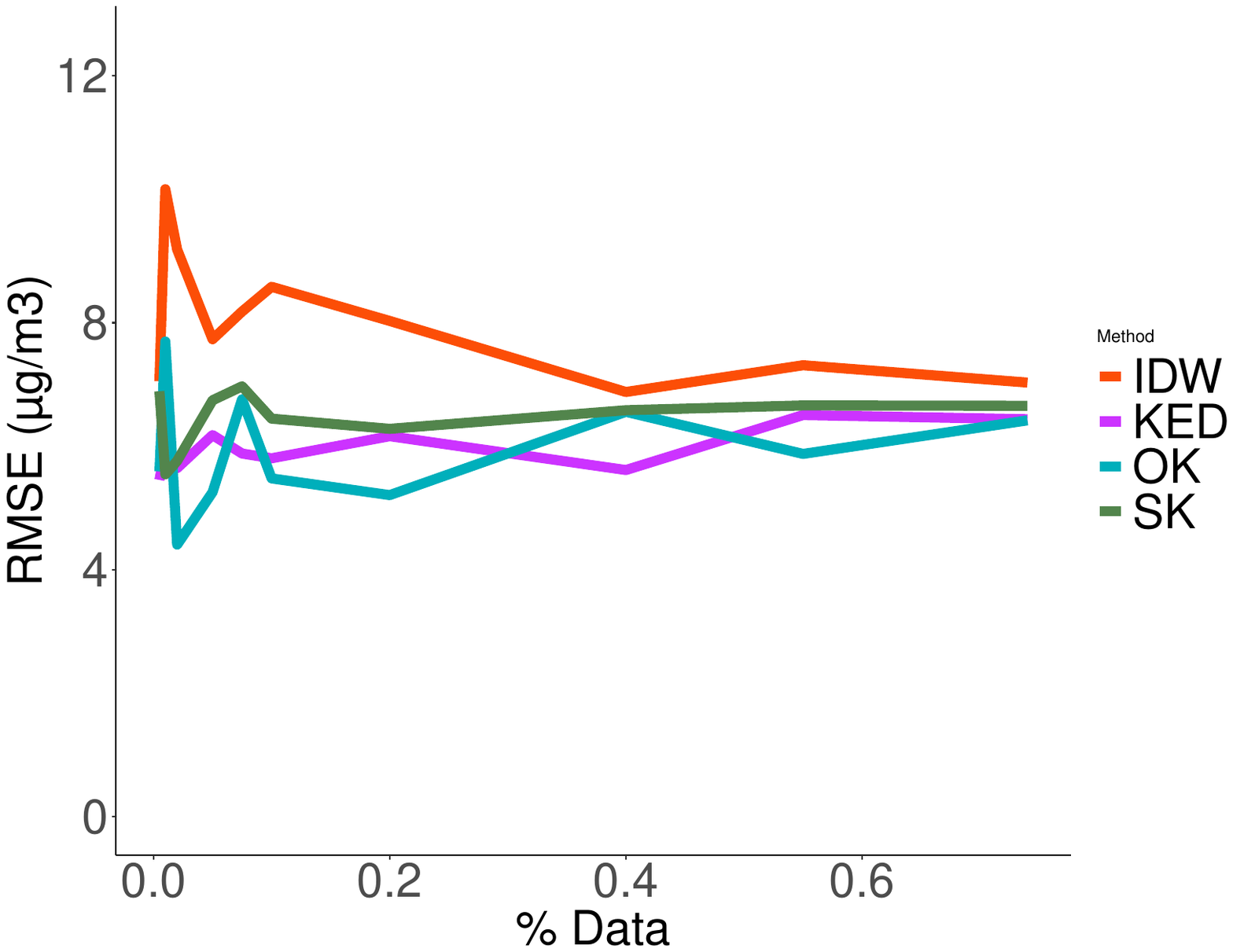} & \includegraphics[width=4.2cm]{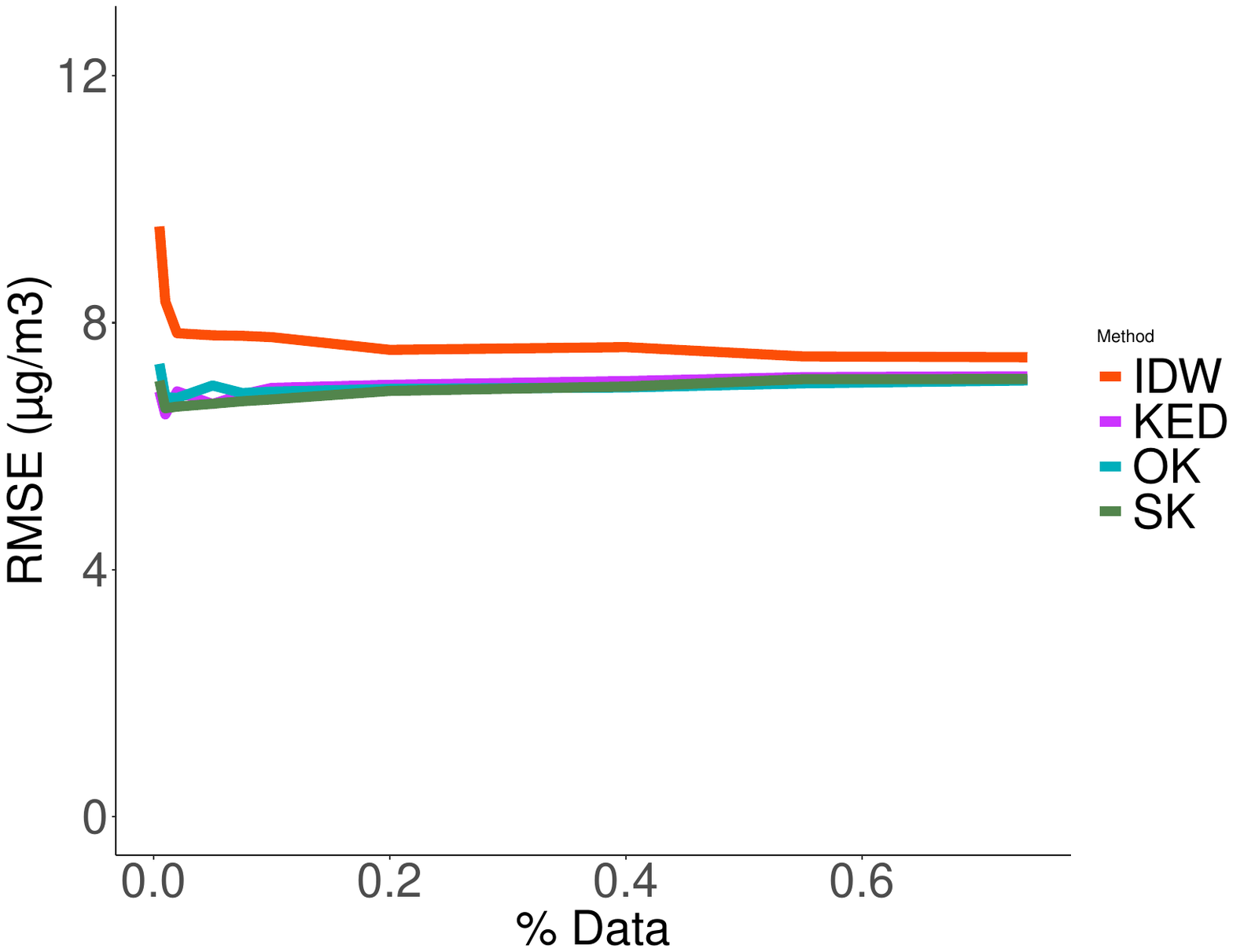} \\
(a) First scenario & (b) Second scenario & (c) Third scenario
\end{tabular}
\caption{RMSE comparison across scenarios.}
\label{RMSE}
\end{figure}

The first thing to notice in the root mean square error (Figure~\ref{RMSE}) is that the three probabilistic methods perform significantly better than the deterministic interpolation in each scenario. As expected, in the first scenario, the more data is used, the less errors are made. This is not true in the third scenario where it is noticed that the error reaches a minimum. No matter how much data is used, the root mean square error does not fall below 6 $\mu$g/m$^3$.

Kriging with external drift shows the least errors in the case of data reconstruction at places close to the sampled data (scenario 1) followed by simple kriging and ordinary kriging. It can be concluded that the contribution of the fixed station data is such environment is useful and that the KED optimizes its use. In the two others scenarios, the use of ordinary kriging seems more appropriate.

\begin{figure}[H]
\centering
\begin{tabular}{ccc}
\includegraphics[width=4.2cm]{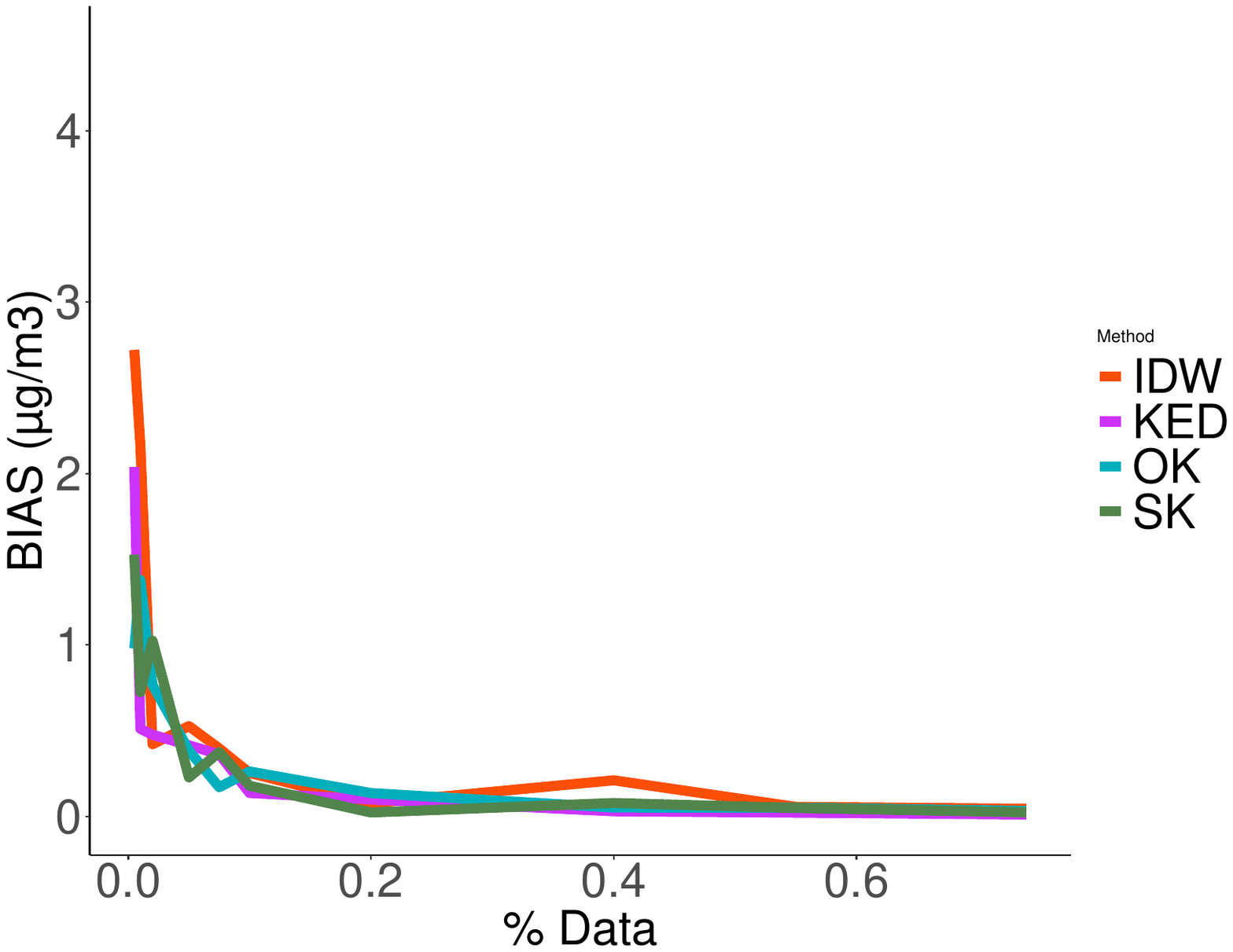} & \includegraphics[width=4.2cm]{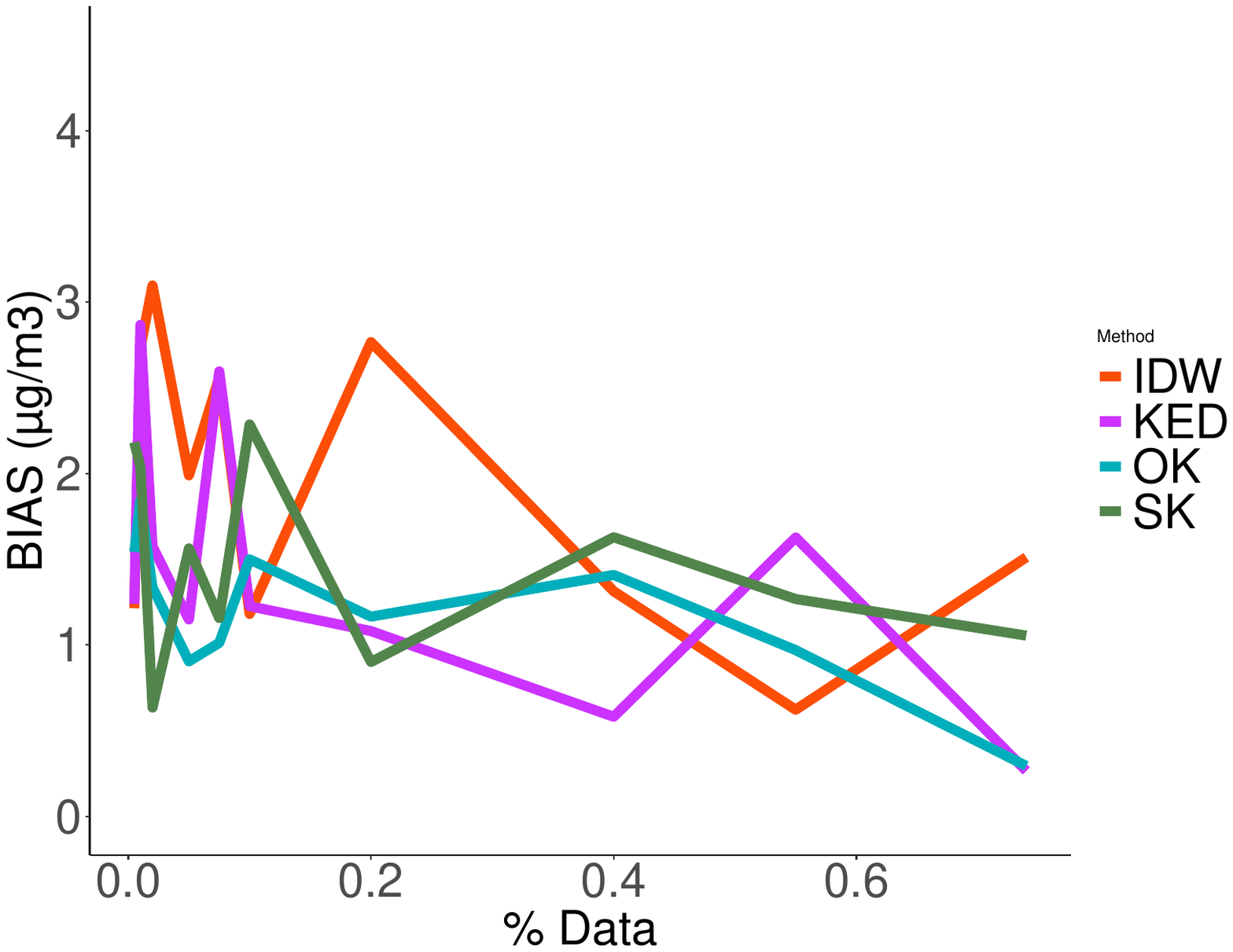} & \includegraphics[width=4.2cm]{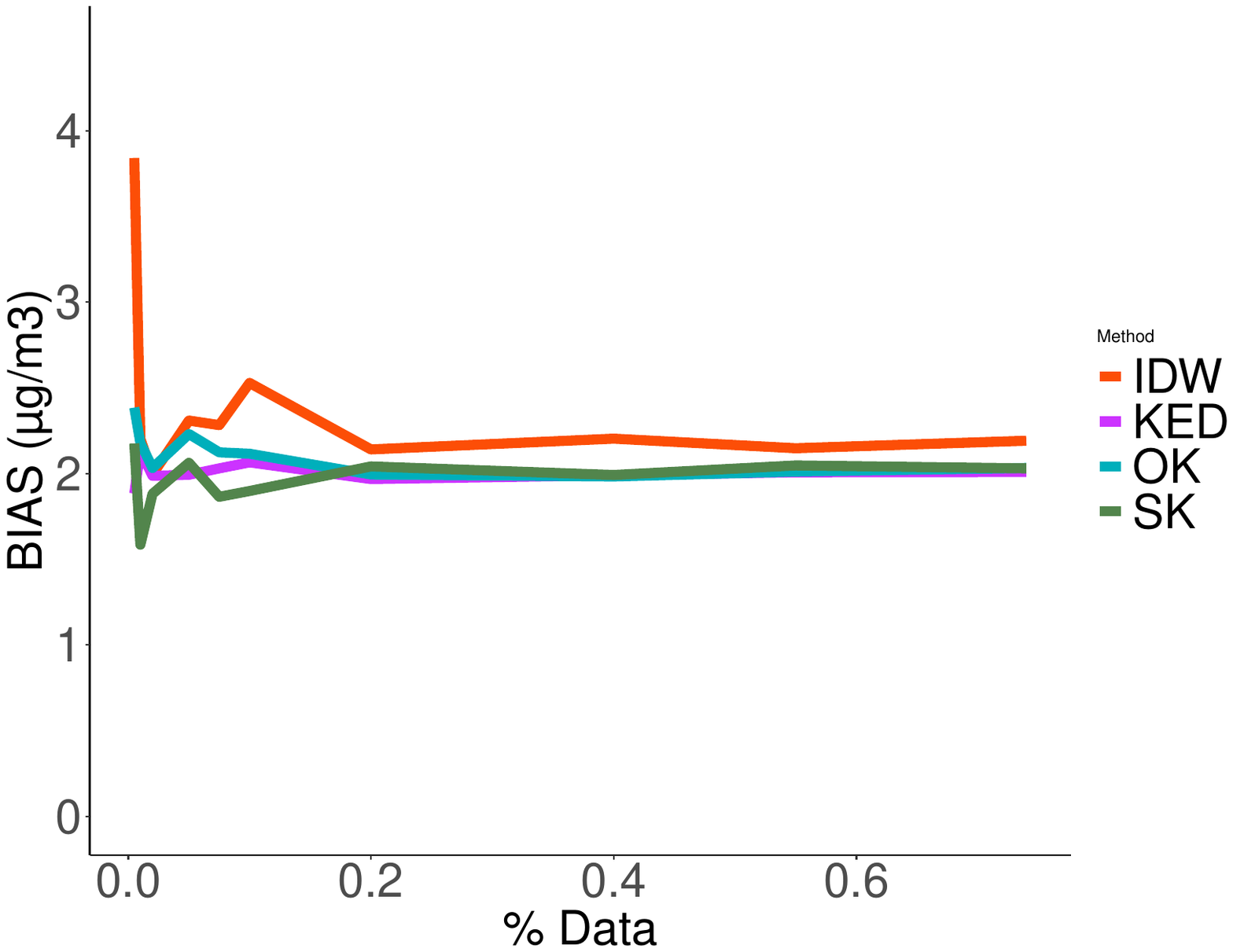} \\
(a) First scenario & (b) Second scenario & (c) Third scenario
\end{tabular}
\caption{BIAS comparison across scenarios.}
\label{BIAIS}
\end{figure}

\begin{figure}[H]
\centering
\begin{tabular}{ccc}
\includegraphics[width=4.2cm]{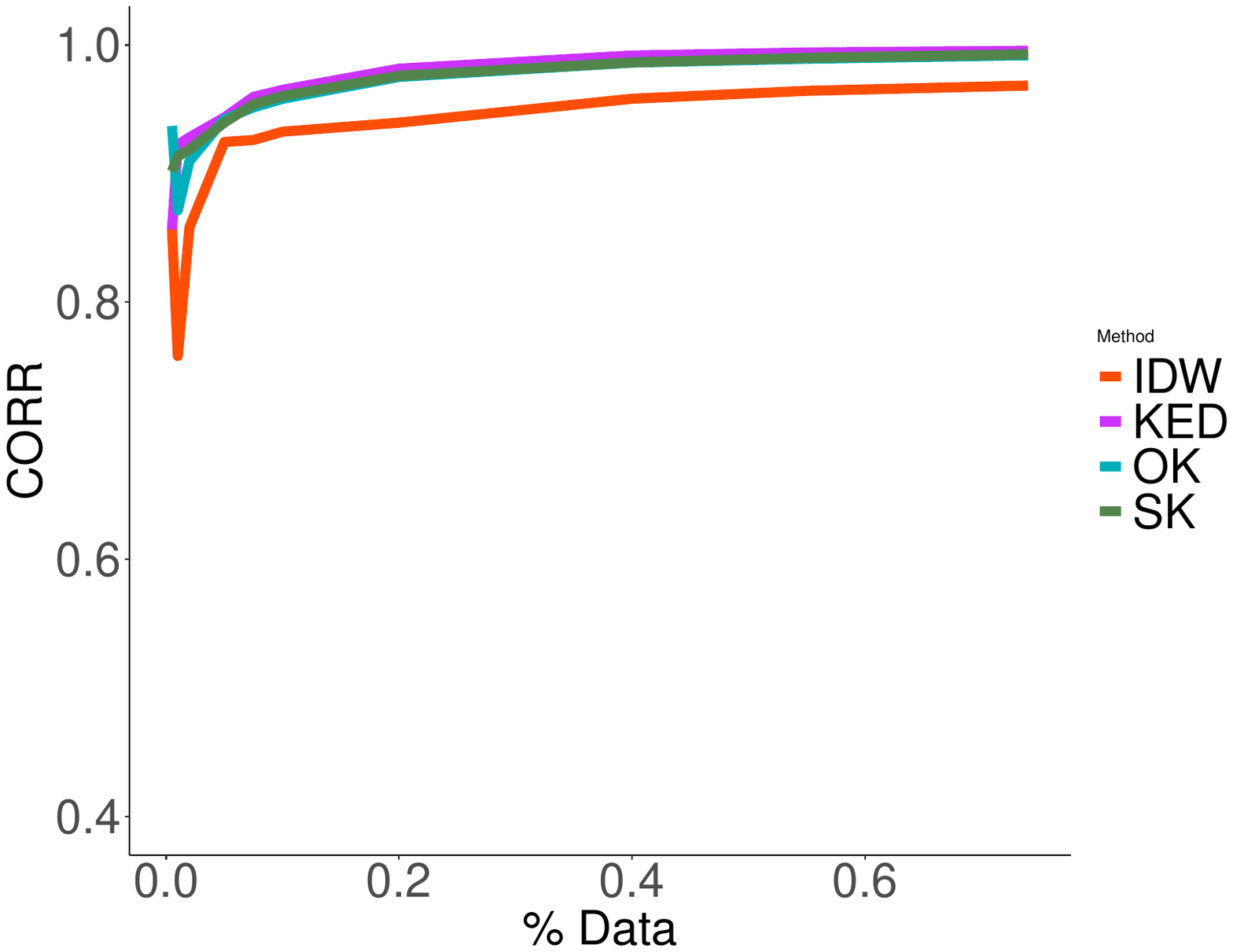} & \includegraphics[width=4.2cm]{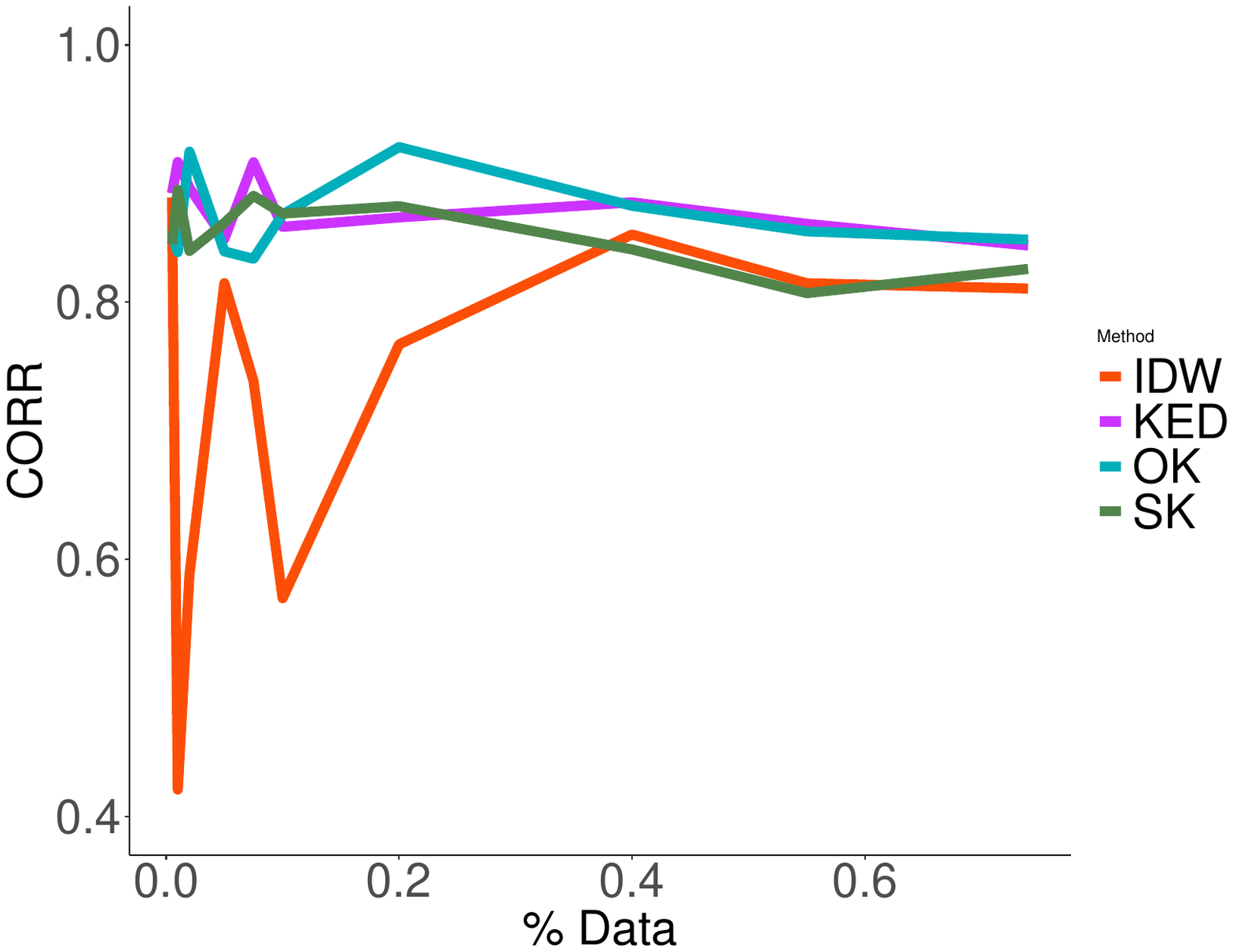} & \includegraphics[width=4.2cm]{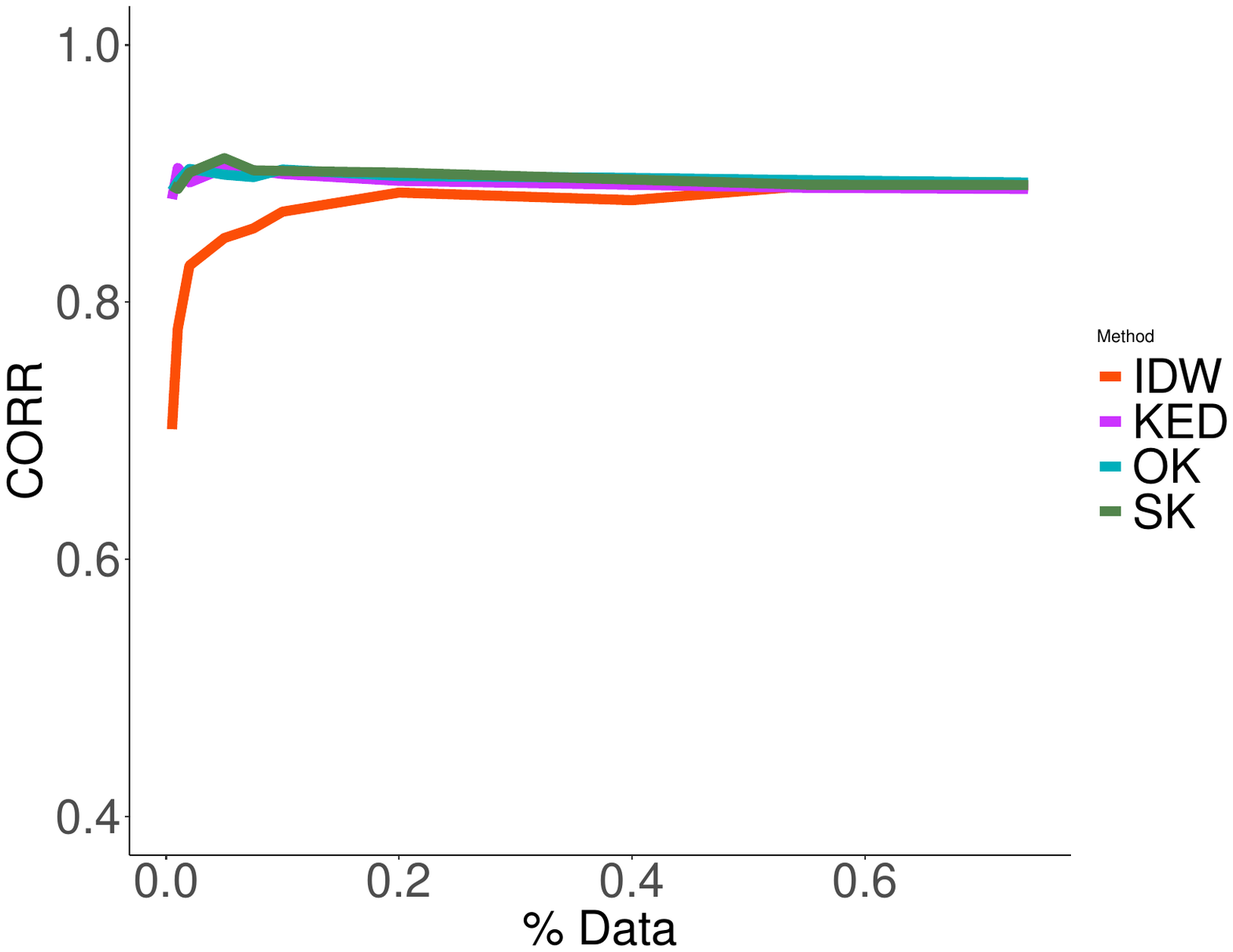} \\
(a) First scenario & (b) Second scenario & (c) Third scenario
\end{tabular}
\caption{CORRELATION comparison across scenarios.}
\label{CORR}
\end{figure}

The four methods bias tend towards zero in the first scenario and in the third scenario the prediction seems systematically biased (Figure~\ref{BIAIS}). Although the stochastic methods systematically outperform the IDW method. This is not the case in the second scenario.
The correlation (Figure~\ref{CORR}) consolidates the idea that kriging with external drift seems best suited in the first scenario, where ordinary kriging shows better correlation results in the second and third scenario.

To summarize, in the first scenario, the performance indicators are smooth, the more data we use, the better we predict. This is not true in the third scenario, where we reach a sill regardless of the number of data points used. As for the second scenario, it is a mix of both.

It can be concluded that kriging using the data from the fixed measurement station as an external variable is the most suitable in the case of data interpolation. When we want to extrapolate far from the sampling places, ordinary kriging seems the good solution.

As expected, and as the majority of data reconstruction methods, geostatistics perform better in the case of interpolation than extrapolation, regardless of the considered performance criterion.

\subsection{Resulting Maps}

\begin{figure}[H]
\centering
\includegraphics[width=\textwidth]{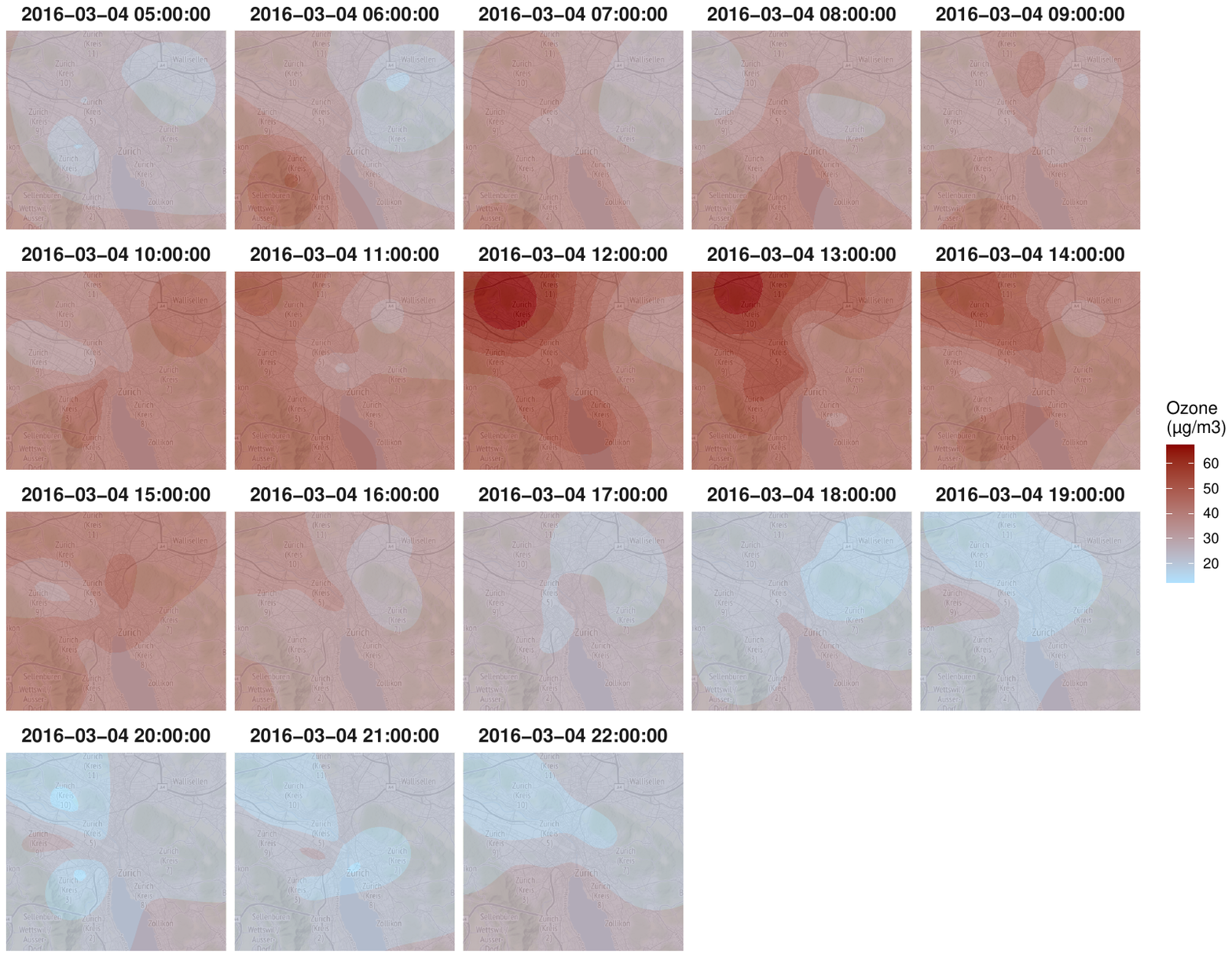}
\caption{Resulting ozone concentrations maps in Zurich from KED algorithm, here shown for 4 March 2016. From 5 am to 10 pm.}
\label{full_plot}
\end{figure}

To answer the objective of creating pollution maps, the KED algorithm was applied using every datapoint available from mobile sensors, as well as the fixed monitoring station during one day. Figure~\ref{full_plot} shows an example of 17 hours of the resulting maps for 4 March 2016.
Figure~\ref{full_plot} displays only the resulting ozone concentration from 5 am to 10 pm, when all four mobile sensors were active.
The method succeeds in identifying areas with high ozone pollution in the city of Zurich, considering that only 4 mobile sensors were used.
One of the important points that can be observed in Figure~\ref{full_plot} is that the typical mid day spike of ozone concentration is clearly visible, followed by mostly very low concentrations during the evening and night.
The concentrations begin to increase throughout the city at around 6 am (depicted by a brief peak observed on lines 13 and 7, as shown by Figures \ref{Scenario1}, \ref{Scenario2} and \ref{Scenario3}).
The concentrations reach a maximum at around 12 am/1 pm, at this point the resulting maps indicate concentrations exceeding 60 $\mu$g/m$^3$ along the north-west side of the city.
Finally, the overall ozone concentration decreases again throughout the evening and around 7 pm reaches approximately the same levels as during the previous night of around 20 $\mu$g/m$^3$ in most areas of the city.

\section{Discussion}

In this paper, several findings need to be highlighted:

\begin{itemize}
\item Spatio-temporal geostatistics offer tools to deal with the problem of using mobile monitoring sensors. While other studies rely on several covariates to predict air quality, this approach can be used to create in real-time air pollution maps. The advantage of geostatistics is that we are not restricted to a given temporal or spatial resolution. Therefore, we can predict at any distance step and any timestep. It would also be possible to predict at greater scales such as road sections or longer time periods using block kriging.
\item Despite the subtraction of the data coming from the fixed stations, there still exist a large spatio-temporal variability which would be easily captured by mobile sensors as it can be seen in the results of this paper.
\end{itemize}

However, several limitations in this study must be detailed:

\begin{itemize}
\item The trams do not go through all types of streets and therefore only measure a specific type of urban pollution. Furthermore, the methodology described above is not used to identify the best model which estimates the ozone concentration, but rather the concentration measured by sensors similar to those used in this study. In this dataset, we do not have access to the real value of the ozone concentration from reference sensors, and it is therefore impossible to carry out a cross validation for this purpose. Moreover, the data from the mobile sensors are considered independent conditionally on ozone concentration and this study doesn't take into account the autocorrelation of data from the same tram.

\item Ordinary kriging does not use the fixed station data in its prediction. Therefore the geostatistical approach can be evaluated in the absence of other data except the ones collected by the mobile sensors. The assumption has been made that the mean is constant but unknown, or at least locally constant being equal to the average of a limited number of datapoints in the neighbourhood of the target point to predict. Thus, this approach is not completely independent from the fixed station data: actually, in the process of sensors calibration using an additive bias (Equation~\ref{eq:1}), the empirical mean of each sensor is imposed to be equal to the mean of the fixed station. Knowing that the ordinary kriging assumes that the average of the field is constant and therefore tends towards the mean of the measurements coming from the mobile sensors, finally the predictions from the ordinary kriging also tend towards the mean of the fixed station.

\item In this study, no model is capable of predicting a value that lies outside of the range of data points on which it is based. Since these interpolations are carried out on subsets of control data, the max and min values in those subsets will be the upper and lower limits of what methods can predict.

\item As no relationship between the spatial coordinates and the variable of interest (ozone concentration) was found in this study, universal kriging couldn't be used. The absence of auxiliary variables makes the prediction outside the collection areas extremely hazardous. As the geostatistics does not create information, one must rely on dependencies with other variables to predict pollutants concentration outside the sampling area.
\end{itemize}

\section{Conclusions and Perspectives}

Air pollution maps with high spatio-temporal precision is of paramount importance and remains an unsolved problem. The use of a mobile sensors fleet, by increasing the spatial coverage, offers a solution to this problem. The use of these devices requires new models to manage these data and produce air quality maps.

In this paper we proposed the study of three spatio-temporal geostatistics methods, and by comparing them to a deterministic interpolation, we concluded that the probabilistic methods systematically outperform the deterministic method. The use of univariable geostatistics gives conclusive results and is more suitable for interpolation at places close to the sampling site. For the extrapolation it will be necessary to use an auxiliary variable in the form of cokriging or regression-kriging.

Despite a higher complexity, the anisotropic models could improve the quality of the prediction. In this paper we only tested a fixed spatial anisotropy in time but another idea would be to search for a possible variation of anisotropy, related for example, to wind and speed directions. Even if univariate geostatistics have its own benefits, future work must assess the added value from using multivariate geostatistics by comparing several methods in term of complexity, error prediction, data used, and so on.

\section*{Author Contributions}

Conceptualization, Y.I. and O.O.; methodology, O.O.; software, Y.I.; validation, V.J., B.S. and P.C.; formal analysis, Y.I.; investigation, Y.I.; resources, O.O.; data curation, Y.I.; writing---original draft preparation, Y.I.; writing---review and editing, O.O.; visualization, O.O.; supervision, V.J.; project administration, B.S.; funding acquisition, P.C.

\section*{Funding}

This research was funded by Gustave Eiffel Université (50\%) and ESTACA (50\%).

\section*{Data Availability}

The dataset used in this study can be found at: \url{https://zenodo.org/record/3355208}

\section*{Conflicts of Interest}

The authors declare no conflict of interest. The funders had no role in the design of the study; in the collection, analyses, or interpretation of data; in the writing of the manuscript, or in the decision to publish the results.

\section*{Abbreviations}

The following abbreviations are used in this manuscript:

\noindent
\begin{tabular}{@{}ll}
IDW & Inverse Distance Weighting \\
SK & Simple Kriging \\
OK & Ordinary Kriging \\
KED & Kriging with External Drift \\
UFP & Ultra Fine Particles \\
ANN & Artificial Neural Network \\
LUR & Land-Use Regression
\end{tabular}

\bibliographystyle{apalike}
\bibliography{bibfile}

\end{document}